\newif\ifjasa
\jasafalse

\ifjasa

\PassOptionsToPackage{unicode}{hyperref}
\PassOptionsToPackage{hyphens}{url}
\PassOptionsToPackage{dvipsnames,svgnames,x11names}{xcolor}
\documentclass[12pt]{article}

\usepackage{tikz}
\usetikzlibrary{graphs}
\usetikzlibrary{arrows.meta}

\usepackage{notations}
\graphicspath{{./figures/}}
\usepackage{algorithm}
\usepackage{algpseudocode}

\algrenewcommand\Require{\State \textbf{Require: }}
\newcommand{\Data}{\State \textbf{Data: }}
\newcommand{\Output}{\State \textbf{Output: }}

\newtheorem{prop}{Proposition}

\newtheorem{lemm}[prop]{Lemma}
\newtheorem{theo}[prop]{Theorem}
\newtheorem{exam}{Example}
\newtheorem{defi}{Definition}
\newtheorem{assu}{Assumption}
\newtheorem{proof}{Proof}

\newtheorem{rema}{Remark}

\usepackage{amsmath,amssymb}

\IfFileExists{upquote.sty}{\usepackage{upquote}}{}
\IfFileExists{microtype.sty}{
  \usepackage[]{microtype}
  \UseMicrotypeSet[protrusion]{basicmath} 
}{}
\makeatletter
\@ifundefined{KOMAClassName}{
  \IfFileExists{parskip.sty}{%
    \usepackage{parskip}
  }{
    \setlength{\parindent}{0pt}
    \setlength{\parskip}{6pt plus 2pt minus 1pt}}
}{
  \KOMAoptions{parskip=half}}
\makeatother
\usepackage{xcolor}
\makeatletter
\ifx\paragraph\undefined\else
  \let\oldparagraph\paragraph
  \renewcommand{\paragraph}{
    \@ifstar
      \xxxParagraphStar
      \xxxParagraphNoStar
  }
  \newcommand{\xxxParagraphStar}[1]{\oldparagraph*{#1}\mbox{}}
  \newcommand{\xxxParagraphNoStar}[1]{\oldparagraph{#1}\mbox{}}
\fi
\ifx\subparagraph\undefined\else
  \let\oldsubparagraph\subparagraph
  \renewcommand{\subparagraph}{
    \@ifstar
      \xxxSubParagraphStar
      \xxxSubParagraphNoStar
  }
  \newcommand{\xxxSubParagraphStar}[1]{\oldsubparagraph*{#1}\mbox{}}
  \newcommand{\xxxSubParagraphNoStar}[1]{\oldsubparagraph{#1}\mbox{}}
\fi
\makeatother

\usepackage{longtable,booktabs,array}
\usepackage{calc} 
\usepackage{etoolbox}
\makeatletter
\patchcmd\longtable{\par}{\if@noskipsec\mbox{}\fi\par}{}{}
\makeatother
\IfFileExists{footnotehyper.sty}{\usepackage{footnotehyper}}{\usepackage{footnote}}
\makesavenoteenv{longtable}
\usepackage{graphicx}
\makeatletter
\def\maxwidth{\ifdim\Gin@nat@width>\linewidth\linewidth\else\Gin@nat@width\fi}
\def\maxheight{\ifdim\Gin@nat@height>\textheight\textheight\else\Gin@nat@height\fi}
\makeatother
\setkeys{Gin}{width=\maxwidth,height=\maxheight,keepaspectratio}
\makeatletter
\def\fps@figure{htbp}
\makeatother

\usepackage[margin=0.8in]{geometry}

\makeatletter
\@ifpackageloaded{caption}{}{\usepackage{caption}}
\AtBeginDocument{%
\ifdefined\contentsname
  \renewcommand*\contentsname{Table of contents}
\else
  \newcommand\contentsname{Table of contents}
\fi
\ifdefined\listfigurename
  \renewcommand*\listfigurename{List of Figures}
\else
  \newcommand\listfigurename{List of Figures}
\fi
\ifdefined\listtablename
  \renewcommand*\listtablename{List of Tables}
\else
  \newcommand\listtablename{List of Tables}
\fi
\ifdefined\figurename
  \renewcommand*\figurename{Figure}
\else
  \newcommand\figurename{Figure}
\fi
\ifdefined\tablename
  \renewcommand*\tablename{Table}
\else
  \newcommand\tablename{Table}
\fi
}
\@ifpackageloaded{float}{}{\usepackage{float}}
\floatstyle{ruled}
\@ifundefined{c@chapter}{\newfloat{codelisting}{h}{lop}}{\newfloat{codelisting}{h}{lop}[chapter]}
\floatname{codelisting}{Listing}

\makeatother
\makeatletter
\@ifpackageloaded{caption}{}{\usepackage{caption}}
\@ifpackageloaded{subcaption}{}{\usepackage{subcaption}}
\makeatother

\newcommand{\anon}{1}

\begin{document}

\def\spacingset#1{\renewcommand{\baselinestretch}%
{#1}\small\normalsize} \spacingset{1}


\if1\anon
{
  \title{\bf Optimal Targeting in Dynamic Systems}
  \author{Yuchen Hu \thanks{
    This research was supported by the Office of Naval Research under grant N00014-24-1-2091.}\hspace{.2cm}\\
    Laboratory for Information \& Decision Systems, MIT\\
    and \\
    Shuangning Li \\
    Booth School of Business, University of Chicago\\
    and \\
    Stefan Wager \\
    Graduate School of Business, Stanford University}
  \maketitle
} \fi

\if0\anon
{
  \bigskip
  \bigskip
  \bigskip
  \begin{center}
    {\LARGE\bf Optimal Targeting in Dynamic Systems}
\end{center}
  \medskip
} \fi

\bigskip
\begin{abstract}
Modern treatment targeting methods often rely on estimating a conditional average treatment effect (CATE) using machine learning tools. While effective in identifying who benefits from treatment on the individual level, these approaches typically overlook system-level dynamics 
that may arise when treatments induce strain on shared capacity. We study the problem of targeting in Markovian systems, where treatment decisions must be made one at a time as units arrive, and early decisions can impact later outcomes through delayed or limited access to resources. 
We show that optimal policies in such settings compare CATE-like quantities to state-specific thresholds, where each threshold reflects the expected cumulative impact on the system of treating an additional individual in the given state.
We propose an algorithm that augments standard CATE estimation with state-level value iteration to estimate these thresholds from observational data. Theoretical results establish consistency and convergence guarantees, and empirical studies demonstrate that our method improves long-run outcomes
considerably relative to individual-level CATE targeting rules and generic offline reinforcement learning algorithms.
\end{abstract}

\noindent%
\vfill

\newpage
\spacingset{1.8} 

\else

\documentclass{article}

\usepackage{amsmath, amsthm, amssymb}
\usepackage{graphicx}
\usepackage{verbatim}
\usepackage{natbib}
\usepackage{caption}
\usepackage{subcaption}
\usepackage{fancyvrb}
\usepackage{enumerate}
\usepackage{relsize}
\usepackage{algorithm}
\usepackage{algpseudocode}
\usepackage{enumitem}
\usepackage{mathtools}
\usepackage{multirow}

\usepackage{hyperref}
\usepackage[margin=1.5in]{geometry}
\hypersetup{colorlinks,citecolor=blue,urlcolor=blue,linkcolor=blue}

\usepackage{tikz}
\allowdisplaybreaks
\usetikzlibrary{graphs}
\usetikzlibrary{arrows.meta}

\usepackage{notations}
\graphicspath{{./figures/}}

\algrenewcommand\Require{\State \textbf{Require: }}
\newcommand{\Data}{\State \textbf{Data: }}
\newcommand{\Output}{\State \textbf{Output: }}


\newcommand{\bZmu}{\widebar{Z\mu}}

\theoremstyle{plain}
\newtheorem{prop}{Proposition}

\newtheorem{conj}[prop]{Conjecture}
\newtheorem{coro}[prop]{Corollary}
\newtheorem{lemm}[prop]{Lemma}
\newtheorem{theo}[prop]{Theorem}

\theoremstyle{definition}
\newtheorem{exam}{Example}
\newtheorem{defi}{Definition}
\newtheorem{assu}{Assumption}
\newtheorem{proo}{Proof}
\newtheorem{model}{Model}

\theoremstyle{remark}
\newtheorem{comm}{Comment}
\newtheorem{rema}{Remark}

\newcommand\blfootnote[1]{%
  \begingroup
  \renewcommand\thefootnote{}\footnote{#1}%
  \addtocounter{footnote}{-1}%
  \endgroup
}

\title{Optimal Targeting in Dynamic Systems}
\author{Yuchen Hu \\ MIT \and
Shuangning Li \\ University of Chicago \and
Stefan Wager \\ Stanford University}

\date{Draft version \ifcase\month\or
January\or February\or March\or April\or May\or June\or
July\or August\or September\or October\or November\or December\fi \ \number%
\year\ \  }


\begin{document}

\maketitle

\begin{abstract}
Modern treatment targeting methods often rely on estimating a conditional average treatment effect (CATE) using machine learning tools. While effective in identifying who benefits from treatment on the individual level, these approaches typically overlook system-level dynamics 
that may arise when treatments induce strain on shared capacity. We study the problem of targeting in Markovian systems, where treatment decisions must be made one at a time as units arrive, and early decisions can impact later outcomes through delayed or limited access to resources. 
We show that optimal policies in such settings compare CATE-like quantities to state-specific thresholds, where each threshold reflects the expected cumulative impact on the system of treating an additional individual in the given state.
We propose an algorithm that augments standard CATE estimation with state-level value iteration to estimate these thresholds from observational data. Theoretical results establish consistency and convergence guarantees, and empirical studies demonstrate that our method improves long-run outcomes
considerably relative to individual-level CATE targeting rules and generic offline reinforcement learning algorithms.
\end{abstract}

\fi

\section{Introduction}

\ifjasa

Flexible machine-learning-based causal inference tools are widely used to guide decision-making and treatment prioritization rules across various domains. The conditional average treatment effect (CATE), which quantifies the expected benefit of treatment conditionally on individual characteristics, serves as a key metric for characterizing treatment heterogeneity and identifying units most likely to benefit from treatment \citep{athey2016recursive}. Many machine learning models have been developed and applied to estimate CATE using rich sets of features. For example, using electronic health records, \citet{zainal2024developing} have found that the causal forest \citep{athey2019generalized} can identify some groups of patients with suicidal ideation who strongly benefit from psychiatric hospitalization, and other groups of patients who may even respond negatively to hospitalization.

\else

Flexible\blfootnote{\hspace{-5mm}This research was supported by the Office of Naval
Research under grant N00014-24-1-2091.}
machine-learning-based causal inference tools are widely used to guide decision-making and treatment prioritization rules across various domains. The conditional average treatment effect (CATE), which quantifies the expected benefit of treatment conditionally on individual characteristics, serves as a key metric for characterizing treatment heterogeneity and identifying units most likely to benefit from treatment \citep{athey2016recursive}. Many machine learning models have been developed and applied to estimate CATE using rich sets of features. For example, using electronic health records, \citet{zainal2024developing} have found that the causal forest algorithm \citep{athey2019generalized} can identify some groups of patients with suicidal ideation who strongly benefit from psychiatric hospitalization, and other groups of patients who may even respond negatively to hospitalization.

\fi

One practical limitation of these machine-learning-based approaches is that they usually focus only on understanding, on a unit-by-unit level, who would benefit from receiving a treatment. However, if an organization such as the Veterans Health Administration wanted to operationalize the findings from these approaches, they would also need to consider how the resulting policy might stress their available resources and infrastructure \citep{hussey2016resources}. In the short-to-medium term, the number of available hospital beds is likely a fixed resource, so beyond a certain point, assigning more patients to inpatient care may cause delays or reduce resource availability for other patients.

The goal of this work is to augment machine-learning-based methods for optimal targeting, such as causal forests, to make them aware of (and responsive to) capacity constraints induced by the system's dynamics that may stochastically limit or delay the number of units that can receive treatment at any given time. We believe this to be a problem setting with broad applicability. For example, such methods could be used for 
targeting expedited delivery in online delivery platforms or
fast-track care in the emergency department. In order to address long wait times in emergency departments, some hospitals implement fast-track systems to divert patients with more minor illnesses from the main intake system and enable them to get faster care \citep{meislin1988fast}. Using a machine learning system to target assignment to the fast-track system could be promising; however, any such system should also be aware of the current queuing conditions and capacity resources of the hospital.

The problem of optimal treatment assignment in a dynamic system is a reinforcement learning problem---and, when working with complex or high-dimensional inputs as are often encountered with, e.g., electronic health records, one might expect this reinforcement learning problem to be intractably complicated. We find, however, our optimal targeting problem admits a simple and intuitive solution: We show that policymakers can account for treatment-induced congestion effects by combining standard treatment prioritization rules with state-dependent decision thresholds.

More specifically, using the hospital admission setting as a motivating example, let $\tau(x,k)$ represent the direct effect of hospitalization for patients with health records $x$ given the level of system congestion $k$; following \cite{munro2025treatment}, we refer to this quantity as Conditional Average Direct Effect (CADE). Then, our results show that the optimal policy must take the form of a CADE-thresholding rule with state-specific thresholds $c_k$, and assigns treatment only to those whose health records $X_i$ satisfy $\tau(X_i,k) > c_k$. 
This structure closely mirrors that of optimal treatment prioritization rules under static budget constraints \citep{bhattacharya2012inferring,sun2021treatment}.
The relationship between these thresholds $c_k$, as well as their magnitude, is shaped by the intricate interplay of the system's dynamics, the distribution of covariates, and the outcome function, each of which can vary largely depending on the context. 
Yet, despite this complexity, we show that these threshold values can be estimated from a single observational dataset by combining standard CADE estimation with a state-level value iteration. In doing so, treatment decisions can be
fine-tuned to strike a balance between individual benefit and system capacity. 

\subsection{Related Work}

The problem of learning optimal targeting rules has been extensively studied over the past two decades, primarily in settings where treatment assignments are considered independently across units \citep{manski2004statistical,manski2009identification,hirano2009asymptotics,stoye2009minimax,stoye2012minimax,qian2011performance,kitagawa2018should,kallus2018balanced,luedtke2020performance,athey2021policy}. Works in this vein have then focused on deriving regret-optimal treatment rules in heterogeneous populations that depend solely on the focal unit's condition. 
In such settings, recent advances in machine learning allow accounting for heterogeneous treatment effects based on a rich set of covariates \citep{chernozhukov2018double,chernozhukov2022locally,athey2019generalized,nie2021quasi,farrell2021deep,kennedy2024minimax}. 
Our approach differs from those by also considering the potential interactions between different treatment assignments that arise from a shared pool of limited resources, particularly in the form of stochastic delay or congestion.

Our work deals with dynamic capacity constraints limiting the number of units eligible for treatment. From this perspective, our research is related to the strand of literature on finding optimal targeting rules under budget constraints \citep{bhattacharya2012inferring,luedtke2016optimal,wang2018learning,le2019batch,huang2020estimating,sun2021empirical,sun2021treatment,xu2022estimating,kitagawa2023should,zhou2023offline,imai2023experimental,sverdrup2024qini,yadlowsky2025evaluating}. However, these works typically assume a fixed, one-time budget---where resources are permanently depleted once allocated---whereas our approach considers dynamic resource capacity, which can recover as units are served and exit the system.

Shared capacity or state can be understood as a form of interference, where the treatment assigned to units arriving earlier may affect the outcomes of units arriving later. 
While there is a growing body of literature that studies the estimation of treatment effects under interference within the Neyman–Rubin framework \citep{hudgens2008toward,tchetgen2012causal,manski2013identification,aronow2017estimating,basse2019randomization,savje2021average,li2022random,leung2022rate,farias2022markovian,zhan2024estimating,johari2024does}, including the recent works that study causal inference and experimentation under stochastic congestion \citep{li2023experimenting,boutilier2024randomized}, few studies have focused on optimal targeting under such conditions. Existing works in this area tend to address the challenge by determining a single targeting rule for the entire population at once \citep{galeotti2020targeting,ananth2020optimal,kitagawa2023individualized,kitagawa2023should,zhang2023individualized,viviano2024policy,munro2025treatment}, whereas the problem we consider involves sequential targeting, where decisions are made one at a time as units arrive in the system. The targeting threshold in our framework is conceptually related to the conditional average indirect effect studied in \cite{munro2025treatment}, which extends the average indirect effect studied in \cite{hu2022average} and \cite{li2022random}. However, unlike those works, our method does not require explicitly estimating this effect through augmented experiments with perturbations.

Our work is also closely related to the literature on off-policy learning in dynamic processes. Early contributions in this field primarily focused on learning a sequence of treatment rules that depend on the entire history of the process \citep{murphy2003optimal,murphy2005generalization,robins2004optimal,zhang2013robust,nie2021learning}. This approach imposes restrictions on both the total number of periods and the complexity of the historical information being tracked. 
In the average reward setting that we consider \citep{howard1960dynamic}, a more common approach is to model the system as a Markov decision process \citep{sutton1999policy,adusumilli2019dynamically,zhang2021policy,kallus2021minimax}. 
The conditional average effects we study are closely related to the difference-of-Q formulation introduced in the recent reinforcement learning literature, which captures the impact of policy deviations \citep{farias2022markovian,cao2024orthogonalized,johari2025estimation}. 
A key distinction between the Markov model we use and those commonly found in the literature is the handling of covariates.
We assume that the units' covariates are exogenous, which allows us to separate individual-level characteristics tied to a unit’s own treatment experience from the system-level state that passes down through the system and affects future units. In contrast, typical MDP models often blur the distinction between individual covariates and system states, thus leading to complex dynamics and inflated policy spaces.

A particularly relevant line of work studies structured Markov decision processes motivated by operational systems and exogenous inputs. One related strand studies mixed systems in queuing-motivated settings, where the state is decomposed into a stochastic component and a pseudo-stochastic component whose update rule is known, which enables virtual sample generation \citep{wei2023mixed}. A separate strand studies exogenous MDP formulations, where the decision maker can exploit known or partially known structure in the endogenous dynamics and rewards while learning the effect of exogenous randomness \citep{dietterich2018discovering,efroni2022sample,kallus2022stateful,sinclair2023hindsight,wan2024exploiting}. These models are related to ours in separating exogenous inputs from endogenous system states, but differ in their statistical focus: in our setting, both the reward function and the state-transition mechanism are unknown and must be learned from data.

Finally, at a higher level, the problem we aim to address is closely related to those studied in the dynamic resource allocation and optimal control literature
\citep[e.g.,][and references therein]{naor1969regulation,johansen1980control,chen2001state,borgs2014optimal,xu2016using,agrawal2019learning,cheung2019sampling,murthy2022admission,chen2024learning,boutilier2024randomized}.  
Much of this literature has focused on managing limited capacity with system-level objectives and often abstracts away individual-level heterogeneity.
More recently, there has been a growing interest in data-driven approaches that leverage covariate or contextual information to improve decision-making
in revenue and inventory management
\citep{feng2018research,ban2019big,bertsimas2020predictive,chen2022statistical,ding2024feature}, as well as in patient prioritization and triage \citep{jacobson2012priority,mills2013resource,shi2024treatment,keskin2024feature,bansak2024dynamic}.
Our work contributes to this literature by introducing a perspective that bridges causal machine learning tools with dynamic constraints, which enables simple, interpretable decision making even in the presence of complex covariates.

\section{Targeting in Dynamic Systems}
\label{sec:general_model}

Consider the problem of targeting in a single dynamic system, where units $i=1,\dots,n$ arrive sequentially. When each unit $i$ arrives, we observe their covariate $X_i \in \mathcal{X}$ and decide on a treatment $W_i \in \cb{0,1}$ to assign. Each time a unit arrives and is assigned a treatment, this may alter the system's dynamics. We capture this dynamic feature using the state variable $S_i \in \mathcal{S}$, which represents the system's condition at the time of the unit's arrival. The unit's outcome, $Y_i \in \mathbb{R}$, may depend on their covariate $X_i$, the treatment assignment $W_i$, and the current state $S_i$. Let $\pi:\mathcal{X}\times \mathcal{S}\to [0,1]$ denote the treatment assignment policy, i.e., 
$\PP{W_i=1\cond X_i=x,S_i=s}=\pi(x,s)$ under policy $\pi$. We use the outcome as a proxy for utility and are interested in learning a good policy $\pi$ from an observed dataset of size $n$ that maximizes the average outcome.

\begin{exam}[Emergency Department Triage]
In a hospital emergency department, patients arrive sequentially, each with observed clinical features $X_i$. The hospital needs to decide whether to offer fast-track treatment ($W_i=1$) or place the patient in the regular queue ($W_i=0$). The system state $S_i$ reflects the numbers of patients currently in each queue. The patient's health outcome may depend on both their clinical features and the queue length of the queue they are assigned to.
\label{exam:triage}
\end{exam}

\begin{exam}[Routing Customer Support Requests]
In an online service platform, customer support tickets arrive sequentially, each with observed features $X_i$ such as account information and request details. The platform decides whether to route the ticket to a specialist ($W_i=1$) or assign it to regular support staff which is not capacity-constrained ($W_i=0$). The system state $S_i$ represents the number of unresolved tickets awaiting specialist attention. The customer’s satisfaction score 
may depend on the assignment, their features, and specialist queue length if routed to them.
\label{exam:support}
\end{exam}

One common targeting strategy is to estimate the Conditional Average Treatment Effect (CATE) from a sample of $n$ units, and implement a policy that targets all units with a positive estimated CATE \citep{manski2004statistical,stoye2012minimax,tetenov2012statistical}. 
In our dynamic setting, the natural state-aware analogue of the CATE is
\begin{equation}
\tau(x,s):=\EE{Y_i\cond X_i=x,S_i=s,W_i=1} - \EE{Y_i\cond X_i=x,S_i=s,W_i=0}.
\end{equation}
However, as we will discuss later, this quantity only captures the direct effect of the treatment on the current individual, and misses its potential impact on later individuals through changes in the system state. Thus, in what follows, we refer to $\tau(x,s)$ as the Conditional Average Direct Effect (CADE), and refer to the CADE-based targeting strategy as the direct targeting rule.

\begin{defi}[Direct Targeting Rule]
\label{defi:CATE_targeting}
Let $\htau(x,s)$ be an estimated CADE from the dataset.
The direct targeting rule $\pi^{\text{d}}$ assigns treatment whenever $\htau(x,s)>0$, i.e., $\pi^{\text{d}}(x,s)=I(\htau(x,s)>0)$.
\end{defi}

One might think that this targeting strategy already accounts for the dynamic nature of the system, since the state variable $S_i$ is also used for targeting. For example, if the treatment represents admission to a system and the state variable represents the busyness level of the system, the direct targeting rule might only assign individuals to the system when it is not busy. However, this argument overlooks how the treatment assignment might affect the system state. Under a direct targeting rule, individuals who benefit only marginally from treatment may still join the system, potentially causing congestion that crowds out individuals for whom treatment would have had substantial benefit.
In fact, as we will show later, the direct targeting rule may only be optimal when the treatment assignment has little impact on state evolution, or when all units are insensitive to the system's state.

We first make the following assumptions about the stability of the system, which allow us to formally define an optimal policy in a dynamic system.

\begin{assu}[Time-Homogeneous MDP with Exogenous Covariates]
\label{assu:MDP}
The system transition is governed by a set of time-homogeneous distributions
$\cb{P_X(\cdot), P_Y(\cdot\cond x, \, s, \, w), P_S(\cdot\cond s, \, w)}$ such that, for all units $i$ and conditionally on the past, $X_i$, $Y_{i}$ and $S_{i+1}$ are drawn according to the densities \smash{$P_X(\cdot)$}, \smash{$P_Y(\cdot\cond X_i, \, S_{i}, \, W_{i})$} and \smash{$P_S(\cdot\cond S_{i}, \, W_{i})$}.
\end{assu}

\begin{assu}[Irreducibility and Boundedness]
\label{assu:stationary}
The process $\cb{S_i\cond i=1,2,\dots}$ induced by any policy $\pi$ is irreducible and aperiodic. Furthermore, the state space $\mathcal{S}$ is finite, and the outcomes $Y_i$ are almost surely uniformly bounded by a constant $M_Y$.
\end{assu}

\begin{figure}[t]
\centering
\resizebox{0.75\linewidth}{!}{
\begin{tikzpicture}
\node (tw1) at (2.75,1) {$\pi$};
\node (ts2) at (4.8,0.4) {$P_S$};
\node (ty1) at (3.8,-1.5) {$P_Y$};
\node[black, draw] (px) at (7.5,-6) {$P_X$};
\node (w0) at (0,2) {Treatment};
\node[black, draw, circle] (w1) at (3,2) {$W_{1}$};
\node[black, draw, circle] (w2) at (6,2) {$W_{2}$};
\node (wc) at (9,2) {$\cdots$};
\node[black, draw, circle] (wt) at (12,2) {$W_{n}$};
\node (s0) at (0,0) {State};
\node[black, draw, circle] (s1) at (3,0) {$S_{1}$};
\node[black, draw, circle] (s2) at (6,0) {$S_{2}$};
\node (sc) at (9,0) {$\cdots$};
\node[black, draw, circle] (st) at (12,0) {$S_{n}$};
\node (y0) at (0,-2) {Outcome};
\node[black, draw, circle] (y1) at (3,-2) {$Y_{1}$};
\node[black, draw, circle] (y2) at (6,-2) {$Y_{2}$};
\node (yc) at (9,-2) {$\cdots$};
\node[black, draw, circle] (yt) at (12,-2) {$Y_{n}$};
\node (x0) at (0,-4) {Covariate};
\node[black, draw] (x1) at (3,-4) {$X_{1}$};
\node[black, draw] (x2) at (6,-4) {$X_{2}$};
\node (xc) at (9,-4) {$\cdots$};
\node[black, draw] (xt) at (12,-4) {$X_{n}$};
\graph {
(s1)->{(s2),(w1)}, (s1)->[dashed] (y1), (w1)->[bend left, dashed] (y1), (w1)->(s2), (x1)->[dashed] (y1), (x1)->[bend left] (w1),
(s2)->{(sc),(w2)}, (s2)->[dashed] (y2), (w2)->[bend left, dashed] (y2), (w2)->(sc), (x2)->[dashed] (y2), (x2)->[bend left] (w2),
(sc)->{(st),(wc)}, (sc)->[dashed] (yc), (wc)->[bend left, dashed] (yc), (wc)->(st), (xc)->[dashed] (yc), (xc)->[bend left] (wc),
(st)->(wt), (st)->[dashed] (yt), (wt)->[bend left, dashed] (yt), (xt)->[dashed] (yt), (xt)->[bend left] (wt),
(px)->(x1), (px)->(x2), (px)->(xc), (px)->(xt) 
};
\end{tikzpicture}
}
\caption{An illustration of a sample of size $n$ drawn from a time-homogeneous Markov decision process with exogenous covariates.
\ifjasa \spacingset{1} \fi
}
\label{fig:mdp}
\end{figure}

A representation of such a Markov decision process with exogenous covariates can be found in Figure \ref{fig:mdp}. 
In Section \ref{sec:queuing}, we provide a concrete example of how this framework naturally models queuing admission and routing systems, and discuss how similar ideas can extend to related dynamic resource-allocation settings.

Assumptions \ref{assu:MDP} and \ref{assu:stationary} ensure that there exists a unique stationary distribution of the state (denoted as $d_{\pi}$), and the average outcome under the policy $\pi$
exists and can be reduced to the following form that does not rely on the initial state:
\begin{equation}
     \lim_{n\to\infty}\EE{\frac{1}{n}\sum_{i=1}^n Y_i \cond S_1=s } = \EE[\pi]{Y_i} =: \mu(\pi),
\end{equation}
where the expectation is taken over the joint distribution of $P_Y$, $P_X$, and the stationary distribution $d_\pi$, i.e., $\EE[\pi]{Y_i} = \int_{\mathcal{X}}\sum_{s\in\mathcal{S}} P_X(x)d_\pi(s)\EE[\pi]{Y_i\cond S_i=s,X_i=x} \operatorname{d}x$ \citep{van1998learning}. We say that a policy is optimal if it maximizes the average outcome $\mu(\pi)$.

\begin{defi}[Optimal Targeting Rule]
    A targeting rule $\pi^*:\mathcal{X}\times \mathcal{S}\to [0,1]$ is optimal if
    \begin{equation}
        \pi^* \in \arg\max_\pi \mu(\pi).
    \end{equation}
\end{defi}

\subsection{A Threshold Characterization with Dynamic Shadow Costs}

A direct targeting rule assigns treatment whenever the current unit's estimated CADE is positive. In a dynamic system, however, this comparison is incomplete. Assigning treatment to the current unit can change the future state of the system, and thus affect the outcomes of later units. 
This raises the question of how the usual direct targeting rule should be adjusted to account for this downstream impact. 
In this section, we show that the adjustment takes a simple form: the optimal policy compares the CADE to a
state-specific threshold $c_s$, where $c_s$ represents the dynamic shadow cost of assigning treatment when the system is in state $s$. 
We first derive this threshold form from the state-level Bellman equation induced by the exogenous-covariate structure, and then connect the characterization to a direct/indirect effect decomposition, where the CADE captures only the direct component of the marginal value of treatment, while our state-specific threshold captures the negative indirect component induced by downstream changes in the system.

For notational convenience, write
\begin{equation}
\eta_w(x,s) := \EE{Y \cond X=x,S=s,W=w},\qquad P_{S,w}(s'|s):= P_S(s'|s,w)
\end{equation}
for the conditional average outcome and the state transition kernel under treatment $w\in\cb{0,1}$. Recall that $\tau(x,s)=\eta_1(x,s)-\eta_0(x,s)$, and write
$\mu^*:= \sup_{\pi}\mu(\pi)$ for the optimal long-run average outcome. Under the exogenous-covariate structure in Assumption~\ref{assu:MDP}, the continuation value of a policy depends only on the system state $s$, rather than on the full covariate-state pair $(x,s)$. 
After integrating over the exogenous covariate distribution, the dynamic component of the problem can therefore be viewed as a finite-state average-reward dynamic program on $\mathcal S$. The standard average-reward Bellman equation for this reduced problem \citep[e.g.,][]{puterman1994markov} then yields the following state-level characterization.

\begin{lemm}
Under Assumptions \ref{assu:MDP} and \ref{assu:stationary},
there exists a relative value function $V^*:\mathcal S\to\mathbb R$ such that
\begin{equation}
\mu^* + V^*(s) = \EE[X]{\max_{w \in \cb{0,1}} {\eta}_w(X,s) +\sum_{s'\in\mathcal S} P_{S,w}(s'|s) V^*(s')},\qquad s\in\mathcal{S}.
\label{eq:bellman_equation_state}
\end{equation}
Moreover, any deterministic policy $\pi^*$ that satisfies
\begin{equation}
\pi^*(x,s) \in \argmax_{w \in \cb{0,1}} \cb{{\eta}_w(X,s) +\sum_{s'\in\mathcal S} P_{S,w}(s'|s) V^*(s')}
\end{equation}
for all $(x,s)\in\mathcal X\times\mathcal S$ is optimal.
\label{lemm:bellman_equation_state}
\end{lemm}

The key feature of the state-level Bellman equation in Lemma~\ref{lemm:bellman_equation_state} is that the relative value function depends only on the state $s$ but not directly on the covariate $x$. 
Below, we show that this separation leads to a particularly simple characterization of the optimal policy. Within each state, the policy compares the CADE to a state-specific continuation-value adjustment. 
Thus, the optimal rule keeps the familiar CADE ranking of the direct targeting rule within each state, but shifts the treatment cutoff according to the continuation-value impact of using capacity in that state.

\begin{theo}
Under Assumptions \ref{assu:MDP} and \ref{assu:stationary},
there exists an optimal deterministic policy of the form
\begin{equation}
    \pi^*(x,s) = I\p{\tau(x,s)> c_s},
\label{eq:policy_continuous}
\end{equation}
where $c_s = - \sum_{s'\in\mathcal{S}} \p{P_{S,1}(s'\mid s)-P_{S,0}(s'\mid s)}V^*(s')$.
\label{theo:policy}
\end{theo}

The threshold $c_s$ can be understood as the dynamic shadow cost of assigning treatment in state $s$. 
Consider the normalized policy gradient at a given covariate-state pair,\footnote{
\ifjasa \spacingset{1}\footnotesize \fi
We divide the policy gradient by $P_X(x)\cdot d_\pi(s)$, the probability of observing the covariate-state pair $(x,s)$, to isolate the conditional impact of adjusting the policy for units with covariate $x$ in state $s$.}
\begin{equation}
H(x,s;\pi) = \frac{1}{P_X(x)\cdot d_\pi(s)}\frac{\partial }{\partial \pi(x,s)} \mu(\pi),
\end{equation}
which represents the marginal value of increasing the treatment assignment probability for units with covariate $X_i=x$ who arrive when the system is in state $S_i=s$. 
By the average-reward policy-gradient theorem \citep{marbach2001simulation}, 
$H(x,s;\pi)$ can be decomposed into the CADE and a Conditional Average Indirect Effect (CAIE):
\begin{equation}
    H(x,s;\pi) = \underbrace{\tau(x,s)}_{\text{CADE}} \, + \, \underbrace{C_{\pi}(s)}_{\text{CAIE}},
\end{equation}
where the CAIE
\begin{equation*}
\begin{split}
C_{\pi}(s) &=  \lim_{n\to\infty}\EE[\pi]{\sum_{i=2}^{n} \p{Y_i-\mu(\pi)}\cond S_1=s,W_1=1}- \lim_{n\to\infty}\EE[\pi]{\sum_{i=2}^{n} \p{Y_i-\mu(\pi)}\cond S_1=s,W_1=0}
\label{eq:indirect}
\end{split}
\end{equation*}
captures the downstream marginal impact of
treatment through future changes in the system state. Under an optimal policy, $c_s=-C_{\pi^*}(s)$.
In other words, $c_s$ is the state-specific dynamic shadow cost of assigning treatment in state $s$ at optimum. For example, in a queuing system, Theorem \ref{theo:policy} suggests that the optimal policy should apply a different treatment threshold at each queue length, and this threshold captures the expected cumulative impact on future outcomes from admitting one additional patient at that queue length under the optimal policy.

We note that our state-aware CADE thresholding approach strictly generalizes the standard direct targeting rule in Definition \ref{defi:CATE_targeting}. 
In the classical setting where the treatment does not affect the system's dynamics, the indirect effect $C_{\pi}(s)$ is always zero.\footnote{\ifjasa \spacingset{1}\footnotesize \fi
This phenomenon, i.e., that the average direct effect of a treatment in a system with spillovers matches the average treatment effect in a system where spillovers are suppressed, has also been observed in many different models \citep[e.g.,][]{savje2021average,munro2025treatment}.} Thus, the optimal policy is simply to compare the CADE with zero, resulting in the direct targeting rule $\pi^*(x,s) = I(\tau(x,s) > 0)$.

\subsection{A State-Aware CADE Thresholding Algorithm}
\label{subsec:OPE_algorithm}

Theorem \ref{theo:policy} suggests that the dynamic targeting problem can be solved by augmenting direct targeting with a state-dependent continuation-value adjustment:
\begin{enumerate}
    \item Estimate $\htau(x, \, s)$ via a standard causal machine learning method, such as a causal forest \citep{athey2019generalized} or an $X$-learner \citep{kunzel2019metalearners}.
    \item Estimate a state-level relative value function $V^*$ by applying relative value iteration~\citep{puterman1994markov} to the empirical Bellman optimality equation, and recover the state-specific treatment thresholds $c_s$.
\end{enumerate}

Writing $r_0(s):=\EE[X]{{\eta}_0(X,s)}$ for the expected reward under treatment $W_i=0$ at state $s$, the state-level Bellman operator can be expressed as
\begin{equation}
\begin{split}
(T_S v)(s) &:= \EE[X]{\max_{w \in \cb{0,1}} {\eta}_w(X,s) +\sum_{s'\in\mathcal{S}} P_{S,w}(s'|s) v(s')}\\
&= r_0(s) + \p{P_{S,0}v}(s) + \EE[X]{\p{\tau(X,s)+ ((P_{S,1}-P_{S,0})v)(s)}_+}.
\end{split}
\label{eq:state_bellman}
\end{equation}
Thus, the Bellman recursion remains state-level, where the covariates enter through the CADE while the continuation-value recursion is defined only on the state space $\mathcal{S}$.

To form an empirical version of the Bellman operator $T_S$, we split the trajectory into a training sample $\mathcal{I}_t$ and an evaluation sample $\mathcal{I}_e$. On the training
sample, we obtain estimates of the CADE, $\htau(x,s)$, and of the transition kernels, $\hat P_{S,w}(s'\cond s)$, $w\in\cb{0,1}$.
The CADE estimate can be obtained using any suitable conditional
treatment effect learner, and the transition kernels can be estimated by empirical conditional transition frequencies
\begin{equation}
\hat P_{S,w}(s'\mid s)
= \frac{\sum_{i\in\mathcal I_t}I(S_i=s,W_i=w,S_{i+1}=s')}
{\sum_{i\in\mathcal I_t}I(S_i=s,W_i=w)}.
\end{equation}
Using the training sample, we also obtain estimates of the conditional outcome $\eta_0(x,s)$ and the behavior policy $\pi_0(x,s):=\PP{W_i=1\cond X_i=x,S_i=s}$,\footnote{We implicitly assume that the status-quo policy depends only on $(X_i, S_i)$. This assumption, often referred to as sequential ignorability \citep{hernan2020whatif,wager2025causal}, is reasonable in our context, as treatment decisions are typically made based on features observable to the system at the time of arrival.} denoted by $\heta_0(x,s)$ and $\hpi_0(x,s)$, for estimating the average baseline outcome $r_0(s)$.

Conditional on these estimates, the
evaluation sample is then used both to estimate the average baseline outcome and to approximate the expectation over $X$ in the state-level
Bellman operator.
Let $\mathcal I_e(s)=\cb{i\in\mathcal I_e:S_i=s}$.
We first estimate $r_0(s)$ through the doubly robust estimator
\begin{equation}
\hat r_0(s) := \frac{1}{\abs{\mathcal{I}_e(s)}} \sum_{i\in \mathcal{I}_e(s)} \sqb{\heta_0(X_i,s) + \frac{I(W_i=0)}{1-\hpi_0(X_i,s)}\cb{Y_i- \heta_0(X_i,s)} }
\label{eq:baseline_estimate}
\end{equation}
using the nuisance components $\heta_0(x,s)$ and $\hpi_0(x,s)$ estimated on the training sample,
and then form the empirical state-level Bellman operator
\begin{equation}
\begin{split}
(\hat T_S v)(s) := \hat r_0(s) + \p{\hat P_{S,0}v}(s) + \frac{1}{\abs{\mathcal{I}_e}}\sum_{i\in \mathcal{I}_e} \p{\htau(X_i,s)+ \p{(\hat P_{S,1}-\hat P_{S,0})v}(s)}_+.
\end{split}
\label{eq:bellman_estimate}
\end{equation}
We summarize our proposed approach to targeting in dynamic systems based on this empirical  state-level Bellman operator in Algorithm~\ref{algo:augmented_cforest_general}.

\begin{algorithm}[t]
\ifjasa \spacingset{1} \fi
\caption{State-Aware CADE Thresholding (SACT) by Relative Value Iteration}
\begin{algorithmic}[1]
\Data $(X_i,S_i,W_i,Y_i)$, $i=1,\dots,n$
\State Split the data into a training sample $\mathcal{I}_t$ and an evaluation sample $\mathcal{I}_e$
\State On the training sample $\mathcal{I}_t$, estimate $\htau(x,s)$, $\hat P_{S,w}(s'\cond s)$, $w=0,1$, and the nuisance functions $\hat\eta_0(x,s)$ and $\hat\pi_0(x,s)$
\State Using $\mathcal I_e$, construct $\hat r_0(s)$ and the empirical Bellman operator $\hat T_S$ as in~\eqref{eq:baseline_estimate} and~\eqref{eq:bellman_estimate}
\State Initialize a value function $V^{(0)}:\mathcal S\to\mathbb R$ and choose an anchor state $s_0$
\For{$m=0,1,2,...$ until convergence}
\State Given $V^{(m)}$, compute $\widetilde{V}^{(m+1)}(s) =(\hat T_SV^{(m)})(s)$ for all $s\in\mathcal S$
\State Recenter with $V^{(m+1)}(s) = \widetilde{V}^{(m+1)}(s) - \widetilde{V}^{(m+1)}(s_0)$, $s\in\mathcal S$,
so that $V^{(m+1)}(s_0)=0$
\EndFor
\Output Set $\hat V^*:=V^{(m+1)}$
and output 
\begin{equation*}
\hpi(x,s) = I\p{\hat{\tau}(x,s)> -\sum_{s'\in\mathcal S}\{\hat P_{S,1}(s'\cond s)-\hat P_{S,0}(s'\cond s)\}\hat V^*(s') }
\end{equation*}
\end{algorithmic}
\label{algo:augmented_cforest_general}
\end{algorithm}

Given a current estimate of the continuation value, Algorithm~\ref{algo:augmented_cforest_general}
selects the treatment action that maximizes the estimated
one-step reward plus continuation value, and then recenters the resulting value function.
Running Algorithm \ref{algo:augmented_cforest_general} requires splitting the sample into a training and evaluation set. This sample-splitting step may initially appear to be less straightforward, as the data points are correlated due to the system’s dynamics. However, for each fixed state $s\in\mathcal{S}$, the system regenerates whenever it returns to state $s$, and the data becomes independent of the preceding trajectory after each regeneration \citep{ross2014introduction}. Therefore, we can split the dataset by first dividing the trajectory into i.i.d. chunks based on occurrences of a chosen state $s$, and then randomly assigning these chunks to the training or evaluation set. 
In practice, one can choose a frequently visited state $s$ to
obtain more balanced regenerative blocks, and use these blocks to implement cross-fitting rather than relying on a single training/evaluation split.

\begin{rema}
When the behavior policy $\pi_0$ is known, the baseline reward $r_0(s)$ can instead be estimated directly by the inverse-probability-weighted estimator
\begin{equation}
\hat r_0^{\mathrm{IPW}}(s)
:=\frac{1}{\abs{\cb{i:S_i=s}}} \sum_{i:S_i=s}\frac{I(W_i=0)}{1-\pi_0(X_i,s)}Y_i .
\end{equation}
The doubly robust form in~\eqref{eq:baseline_estimate} is used to accommodate the more general setting where the behavior policy is unknown and may need to be estimated, possibly with high-dimensional covariates. In this case, directly plugging in
$\hat\pi_0$ in the IPW estimator can introduce first-order sensitivity to behavior-policy estimation error. The augmentation by the baseline outcome regression $\hat\eta_0$ removes this first-order sensitivity, so that the remaining nuisance bias is of the usual product form.
\end{rema}




\subsection{Statistical Guarantees}

We next provide some generic upper bounds---under abstract high-level conditions---on the regret attained by our proposed procedure. For a stationary policy $\pi: \mathcal X \times \mathcal S \to \cb{0,1}$, define
\begin{equation}
P_{S,\pi}(s'\cond s):=\EE[X]{\pi(X,s) P_{S,1}(s'|s) + (1-\pi(X,s)) P_{S,0}(s'|s)},
\end{equation}
and similarly
\begin{equation}
\hat P_{S,\pi}(s'|s)
:= \frac{1}{\abs{\mathcal{I}_e}}\sum_{i\in\mathcal{I}_e}\p{\pi(X_i,s)\hat P_{S,1}(s'|s) + (1-\pi(X_i,s))\hat P_{S,0}(s'|s)}.
\end{equation}
We impose the following uniform mixing assumption on both the true and the empirical transition kernels.

\begin{assu}[Uniform Mixing]
There exist constants $t_0\ge 0$ and $t_{\mix}\ge 1$ such that, for all $t \ge t_0$ and every stationary policy $\pi$, the kernels $ P_{S,\pi}$ and $\hat P_{S,\pi}$ admit unique stationary distribution $d_{\pi}$ and $d_{\hat P, \pi}$, and
\begin{equation}
\sup_{s \in \mathcal S}
\abs{P_{S,\pi}^t(\cdot\cond s) - d_{\pi}}_{\mathrm{TV}}
\le
e^{-t/t_\mix},
\quad
\sup_{s \in \mathcal S}
\abs{\hat P_{S,\pi}^t(\cdot\cond s) - d_{\hat P, \pi}}_{\mathrm{TV}}
\le
e^{-t/t_\mix}.
\end{equation}
\label{assu:uniform_mixing}
\end{assu}

Under Assumptions~\ref{assu:MDP}-\ref{assu:stationary}, the true kernel $P_{S,\pi}$ is induced by a finite-state irreducible and
aperiodic Markov chain, and hence such a finite mixing
time always exists. 
By contrast, the empirical kernel $\hat P_{S,\pi}$ may not automatically inherit these properties. In practice, an anchored version of $\hat P_{S,\pi}$ can often be used to ensure that the above mixing assumption is satisfied by mixing $\hat P_{S,\pi}$ with a fixed reference kernel; see, for example, \citet{zurek2024plug}.

\begin{lemm}
Suppose Assumptions~\ref{assu:MDP}-\ref{assu:uniform_mixing} hold. Then, 
\begin{equation}
\begin{split}
\Norm{\hat V^*-V^*}_\infty
&\lesssim \p{t_0 + t_{\mix}} \p{\Norm{\hat r_0-r_0}_\infty + \max_{s\in\mathcal{S}}\Norm{\htau(\cdot,s)-\tau(\cdot,s)}_{L_2(P_X)} } \\
&\qquad\qquad+ M_Y \p{t_0 + t_{\mix}}^2 \max_{w\in\cb{0,1}}\Norm{\hat P_{S,w}-P_{S,w}}_{\infty,1}+\oop\p{\abs{\mathcal{I}_e}^{-1/2}}.
\end{split}
\end{equation}
where
$
\Norm{\hat P_{S,w}-P_{S,w}}_{\infty,1}
:=
\max_{s \in \mathcal S}
\sum_{s'\in \mathcal S} \abs{\hat P_{S,w}(s' \cond s)-P_{S,w}(s' \cond s)}.
$
\label{lemm:V_error}
\end{lemm}

Lemma~\ref{lemm:V_error} shows how estimation error propagates through the dynamic system under the exogenous-covariate formulation. Errors in estimating the baseline reward and the CADE create local Bellman errors whose impact accumulates over future periods at a rate proportional to the mixing time. By contrast, errors in estimating the transition kernels enter through the continuation-value term and are therefore amplified by both the mixing time and the size of the downstream value function, which leads to the stronger $t_\mix^2$ dependence. 
This decomposition is reminiscent of the simulation lemma for discounted MDPs
\citep{kearns2002near,agarwal2026rlt}, in which estimation error in the immediate reward component propagates through the system linearly in the stability factor, whereas transition-kernel error is amplified by an additional factor because it acts through the downstream value function.

In particular, the potentially complex covariate $X$ affects the bound only through estimation of the CADE $\tau(x,s)$ and does not enter recursively through a value function on the full $(X,S)$ space.
This separation is especially helpful because CADE estimation can often be handled with modern doubly robust or orthogonal methods, whereas the dynamic burden is confined to the much smaller state-level transition object. In Section~\ref{sec:comparison}, we discuss how this separation can substantially reduce the statistical complexity of standard plug-in policy-learning methods in a tabular setting.

Before stating the regret bound, we impose a standard nuisance condition for the doubly robust baseline reward estimator $\hat r_0(s)$. This condition requires overlap for the behavior policy and the usual product-rate control for the nuisance estimates. The additional boundedness requirements on the estimated nuisance functions are technical stability conditions, and can be enforced in practice by truncating $\hpi_0$ away from zero and one and clipping $\heta_0$ to the bounded outcome range.

\begin{assu}[Overlap and Nuisance Regularity]
The behavior policy satisfies the overlap condition that there exists a constant $\Gamma_1>0$ such that $\Gamma_1 \le \pi_0(x,s)\le 1-\Gamma_1$
for all $(x,s)\in \mathcal{X}\times\mathcal{S}$.
The nuisance estimates $\hat\eta_0(x,s)$ and $\hat\pi_0(x,s)$ satisfy
\begin{equation}
    \max_{s\in\mathcal{S}}\Norm{\hat\eta_0(X_i,s)-\eta_0(X_i,s)}_{L_2(P_X)}\cdot \Norm{\hat\pi_0(X_i,s)-\pi_0(X_i,s)}_{L_2(P_X)} =\oo_p\p{\frac{1}{\sqrt{n}}}.
\end{equation}
Moreover, $\hpi_0(x,s)$ is bounded away from zero and one and $\heta_0(x,s)$ is uniformly bounded almost surely for all $(x,s)$.
\label{assu:r0_nuisance}   
\end{assu}

\begin{theo}
Under Assumptions~\ref{assu:MDP}-\ref{assu:r0_nuisance}, if $\htau(x,s)$ learned on the training set $\mathcal{I}_t$ satisfies
\begin{equation}
    \max_{s\in\mathcal{S}}\Norm{\htau(X_i,s)-\tau(X_i,s)}_{L_2(P_X)} = \oo_p\p{\abs{\mathcal{I}_t}^{-\beta}}, \ \ \beta > 0,
\label{eq:tau_rate}
\end{equation}
then the policy $\hpi$ generated by Algorithm~\ref{algo:augmented_cforest_general} with $\abs{\mathcal{I}_t}, \abs{\mathcal{I}_e} \asymp n$ satisfies
\begin{equation}
\begin{split}
\mu(\pi^*) - \mu(\hpi)  &= \oop\p{n^{-(\beta \land \frac{1}{2})}}.
\end{split}
\label{eq:rate_policy}
\end{equation}
\label{theo:policy_consistency}
\end{theo}

Theorem~\ref{theo:policy_consistency} shows that augmenting the direct CADE-thresholding rule
with state-aware thresholds yields regret of order $\oop(n^{-(\beta \land \frac{1}{2})})$. 
In flexible nonparametric settings, the CADE estimation rate is typically no faster than the parametric rate, so the leading term in this bound is often the error in estimating $\tau(x,s)$. 
In this sense, our result has the same qualitative flavor as plug-in thresholding results in the i.i.d. setting, where a basic bound controls regret by the error in
estimating the conditional effect, while sharper plug-in rates typically require additional margin assumptions controlling the amount of covariate mass near the
decision boundary \citep{luedtke2020performance}.
In our setting, the analogous margin condition would control the amount of covariate mass near the state-specific decision boundaries $\tau(x,s)=c_s$, but we do not impose such conditions here.
We also note that our assumed uniform $L_2(P_X)$-norm bound on the estimated CADE function $\htau(X_i,s)$ over states $s\in\mathcal{S}$, i.e., \eqref{eq:tau_rate}, can be attained when the state space $\mathcal{S}$ is finite (as in Assumption \ref{assu:stationary}) by leveraging a large body of literature on non-parametric CATE estimation \citep[see, e.g.,][for a discussion]{kennedy2023towards}.\footnote{\ifjasa\spacingset{1}\footnotesize\fi 
One potential concern is that the units in our setting are correlated, and thus theories developed for CATE estimation in i.i.d. settings may not directly apply. However, we note that it is possible to construct a subset of the data with size $\oo_p\p{n}$ by randomly selecting a single data point from each i.i.d. chunk, formed based on occurrences of some fixed chosen state $s$. This approach allows us to retain i.i.d.-like properties, so we still expect the same convergence rate to hold.} 

\subsection{Comparison with the General Collapsed-State Approach}
\label{sec:comparison}

A key idea of our approach has been our ability to separate individual characteristics from
the system state: We assume that the characteristics $X$ capture heterogeneity in the current unit's outcomes, while the dynamic consequences of current decisions are carried forward by a lower-dimensional system state $S$. 
This separation allows the Bellman recursion in Section~\ref{subsec:OPE_algorithm} to remain lower dimensional.
To elucidate the value of this structure, it is natural to ask what would happen if instead we ran an algorithm that ignores this separation and treated the pair $(X,S)$ as a single Markov state, as one would if tackling our problem with off-the-shelf reinforcement learning tools. Does our proposed approach out-perform such an off-the-shelf baseline---and, if yes, by how much?


When the covariate and the system state are collapsed into a single Markov state, $Z := (X,S) \in \mathcal X \times \mathcal S$, the problem becomes a standard Markov decision process with state variable $Z$, and the corresponding Bellman operator takes the form
\begin{equation}
(Tv)(z) := \max_{w \in \cb{0,1}} 
\p{\eta_w(z) + \sum_{z' \in \mathcal X \times \mathcal S}P_{Z,w}(z'\cond z) v(z')},
\label{eq:poisson}
\end{equation}
For simplicity, we here focus the tabular setting where both $\mathcal X$ and $\mathcal S$ are finite, so relevant model primitives can be estimated by plug-in sample averages.
This enables us to compare our proposed method to standard methods for tabular MDPs \citep[e.g.,][]{kearns2002near}, which provide a direct benchmark for assessing the statistical benefit
of treating $(X,S)$ as a generic Markov state rather than exploiting the exogenous-covariate
structure.


For our proposed method, we form a plug-in version of the state-level operator~\eqref{eq:state_bellman} by estimating $\tau(x,s)$, $r_0(s)$, the covariate
distribution $P_X$, and the state-level transition kernels $P_{S,w}(\cdot\cond s)$ via tabular averaging from the data, and then applying
relative value iteration on the state space $\mathcal S$. 
For the collapsed-state baseline, we instead estimate the reward functions $\eta_w(z)$ and the collapsed transition kernels $P_{Z,w}(\cdot\cond z)$ on the full state space $\mathcal X\times\mathcal S$, and apply relative value iteration to the Bellman operator~\eqref{eq:poisson}.
The following result provides regret bounds for both approaches via the same standard proof technique, i.e., by using concentration bounds for plug-in estimates to control the Bellman-operator error and then propagating this error through the Bellman equation, as in standard simulation-lemma analyses of model-based MDPs \citep{agarwal2026rlt}.

\begin{prop}
Under Assumptions~\ref{assu:MDP}-\ref{assu:r0_nuisance}, suppose in addition that $\mathcal X$ is finite and that there exists a constant $\Gamma_2 >0$ such that, for all $(x,s)\in\mathcal X\times\mathcal S$,
\begin{equation}
P_X(x)\, d_{\pi_0}(s)\ge \frac{\Gamma_2}{|\mathcal X||\mathcal S|}.
\label{eq:overlap_collapsed}
\end{equation}
Then, our method $\hat\pi$ and the collapsed-state baseline $\hat\pi^{\col}$ have regret bounded as
\begin{equation}
\begin{split}
&\sqrt{n}\p{\mu(\pi^*)-\mu(\hat\pi^\col)}
= \oo_p\p{\p{t_0 + t_{\mix}}^2 |\mathcal X||\mathcal S|\sqrt{\log(|\mathcal X||\mathcal S|)} }, \\
&\sqrt{n}\p{\mu(\pi^*)-\mu(\hat\pi)}=\oop\p{\p{t_0 + t_{\mix}}\,\sqrt{|\mathcal X||\mathcal S|\log(|\mathcal X||\mathcal S|)}+ \p{t_0 + t_{\mix}}^2\,|\mathcal S|\sqrt{\log(|\mathcal S|)}}.
\end{split}
\label{eq:regret_cmp}
\end{equation}
\label{prop:collapsed_state_regret}
\end{prop}

Comparing the scaled regret terms on the right-hand-side of \eqref{eq:regret_cmp} we see that $\hat{\pi}$, i.e., the tabular version of our proposed method that separates exogenous covariates from system state, outperforms the collapsed-state baseline by an order of magnitude. Relative to the baseline (and ignoring log factors), the regret bound for $\hat{\pi}$ is reduced by a factor of $\min\{\p{t_0 + t_{\mix}} \sqrt{|\mathcal X||\mathcal S|}, \, |\mathcal X|\}$. This highlights how, when the covariate space $|\mathcal X|$ is even moderately large, our approach is able to achieve considerably stronger guarantees by exploiting the model structure.

Now, of course, Proposition \ref{prop:collapsed_state_regret} doesn't give a full picture on the relative merits of our approach relative to collapsed-state baselines. First, the result only compares regret upper bounds to each other---although we do emphasize that the upper bounds were derived using the same (standard) technique of plug-in Bellman-operator perturbation. Second, this result only compares various implementations of tabular value iteration. A full analysis of minimax regret performance for optimal targeting in our model with exogenous covariates would be of considerable interest, but is beyond the scope of this paper. In our numerical experiments, we will also compare our approach to different collapsed-state reinforcement learning algorithms, and find that our approach outperforms these baselines in experiments.


\section{State-Dependent Arrivals}
\label{subsec:reward_rate}

So far, we have treated the process as indexed by arriving decision units and evaluated a policy by the average outcome per observed unit. This objective is natural when the arrival process is exogenous to the system state and the targeting policy. In some systems, however, the policy can affect not only the outcomes of individuals who enter the system, but also the frequency with which such individuals arrive.
For example, users may choose to leave or avoid a digital platform if they anticipate long waiting times. 
Consequently, the objective may shift from maximizing the average outcome per unit to optimizing the overall benefit accumulated over a fixed time period, such as daily revenue.

To accommodate this setting, we consider an event-level representation that records the timing of events in calendar time.
For concreteness, suppose the system is an $M_n/M/1$ queue with state-dependent arrivals.
In an $M_n$/$M$/1 queue, units arrive at rate
$\lambda_k$ when the current queue length is $k$, and service completions occur at rate $\mu$.
We observe a sequence of events $\{\mathbf{E}_i\}_{i=1,\dots,N(T)}$ occurring at times
$0<T_1<T_2<\dots<T_{N(T)}<T$, where
$N(t) = \max\cb{i:T_i\le t}$ counts the number of events up to time $t$. Each event is either an arrival, denoted by $A_i=1$, or a service completion, denoted by $A_i=0$.

If event $i$ is an arrival, we observe the individual's covariates
$X_i\in\mathcal{X}$ and the current queue length $K_i=0,1,\dots,\bk$, assign a treatment $W_i\sim\pi(X_i,K_i)$ indicating whether to admit the individual into the queue,
and record the reward $R_i$. 
If event $i$ is a service completion, no treatment decision is made, and we set $X_i=\emptyset$, $W_i=2$, and $R_i=0$ for bookkeeping. 
Let $\Delta_i := T_i - T_{i-1}$ denote the inter-event time. Since both the arrival rate and the event type depend on the current queue length, $\Delta_i$ is generally correlated with $K_i$.

To summarize, we write $\mathbf{E}_i = \p{X_i,K_i,A_i,W_i,\Delta_i,R_i}$, where $X_i$ is the incoming unit's condition, $A_i$ is the event type, $K_i$ is the queue length at time $T_i$, $\Delta_i$ is the inter-event time, $W_i$ is the treatment assigned to the incoming unit, and $R_i$ is the unit's outcome. 
This event-level representation has the same Markov structure as the model in Section~\ref{sec:general_model}. In particular, for any fixed scalar outcome constructed from $(R_i,\Delta_i)$, such as the transformed outcome introduced below, the state-aware Bellman characterization can be applied with state $S_i=(K_i,A_i)$.

The framework in Section \ref{sec:general_model} optimizes an average outcome over the indexed process.
In the present setting, however, one may instead be interested in maximizing the long-run reward rate
\begin{equation}
    \theta(\pi) = \EE[\pi]{\lim_{T\to\infty}\frac{1}{T}\sum_{i=1}^{N(T)}R_i }.
\end{equation}
In other words, $\theta(\pi)$ measures the average reward the system achieves per unit of time in the long run.
Such average-reward-per-unit-time objectives are commonly studied in Markov
renewal and semi-Markov decision processes because they account for the
policy-dependent frequency with which rewards are realized
\citep{puterman1994markov,ross2014introduction}.
When arrivals are exogenous and occur at a policy-independent rate, maximizing the average reward per arrival is equivalent to maximizing the reward rate. When the arrival rate varies with congestion, however, the two objectives can differ.
For example, if the treatment is in terms of healthcare delivery, one would optimize $\mu(\pi)$ if the goal is to improve health outcomes on average for every patient that arrives at the hospital. On the other hand, one would optimize $\theta(\pi)$ if the goal is to maximize the total benefit (e.g., the overall amount of health improvement) obtained within a fixed amount of time. 

A standard result for Markov renewal processes (see, e.g., Chapter 7 of \cite{ross2014introduction}) implies that, for any fixed policy $\pi$, the long-run reward rate can be written as
\begin{equation}
\theta(\pi) = \EE[\pi]{R_i }/ \EE[\pi]{\Delta_i },
\end{equation}
where the expectations are taken under the stationary distribution of the event-level process induced by $\pi$.
In other words, the long-run reward rate can be written as the ratio of the average reward per unit, $\EE[\pi]{R_i}$, to the average inter-event time, $\EE[\pi]{\Delta_i}$.
Thus, optimizing the long-run reward rate could be regarded as a fractional optimization problem.

This ratio representation suggests a simple way to reuse the state-level Bellman characterization from Section~\ref{sec:general_model}. Let
\begin{equation}
\theta^* := \max_\pi \theta(\pi) = \max_\pi\frac{\EE[\pi]{R_i}}{\EE[\pi]{\Delta_i}}
\end{equation}
denote the optimal reward rate.
Consider the following transformed outcome $\widetilde Y_i = R_i - \theta^* \Delta_i$.
For any policy $\pi$,
\begin{equation}
\EE[\pi]{\widetilde Y_i} = \EE[\pi]{R_i} - \theta^*\EE[\pi]{\Delta_i}
=\EE[\pi]{\Delta_i}\p{\theta(\pi) - \theta^* }\le 0,
\end{equation}
where the inequality follows from $\EE[\pi]{\Delta_i}>0$ and
$\theta(\pi)\le \theta^*$, and equality holds if and only if
\(\theta(\pi)=\theta^*\).
Therefore, the optimal transformed average outcome is zero, and the policies maximizing this transformed average outcome are exactly the reward-rate optimal policies.

We can thus directly apply the state-level Bellman characterization from Section~\ref{subsec:OPE_algorithm} to the transformed outcome $\widetilde Y_i$. 
In particular, this amounts to replacing the
baseline reward and direct effect in~\eqref{eq:state_bellman} by
$
r_{\widetilde Y,0}(s)
= r_{R,0}(s)-\theta^*r_{\Delta,0}(s)$ and $\tau_{\widetilde Y}(x,s)
= \tau_R(x,s)-\theta^*\tau_\Delta(x,s)$,
where $r_{R,0}(s)$, $r_{\Delta,0}(s)$, $\tau_R(x,s)$, $\tau_\Delta(x,s)$ are the state-level average baseline outcomes and CADEs associated with the two outcome variables $R_i$ and $\Delta_i$. The Bellman operator now becomes
\begin{equation}
(T_{S,\theta^*} v)(s) := r_{\widetilde Y,0}(s) + \p{P_{S,0}v}(s) + \EE[X]{\p{\tau_{\widetilde Y}(X,s)+ ((P_{S,1}-P_{S,0})v)(s)}_+},
\label{eq:state_bellman_theta}
\end{equation}
and an optimal reward-rate policy then takes the form
\begin{equation}
\pi^*(x,s)=I\p{\tau_{\widetilde Y}(x,s) > -\sum_{s'\in\mathcal S}\{ P_{S,1}(s'\cond s)- P_{S,0}(s'\cond s)\} V^*(s') }.
\end{equation}

In practice, since the
Bellman operator $T_{S,\theta^*}$ depends on the unknown parameter
$\theta^*$, we solve the transformed problem by
iterating over candidate reward rates.
Given a current value $\theta^{(b)}$, we apply Algorithm~\ref{algo:augmented_cforest_general} with the Bellman operator $T_{S,\theta^{(b)}}$ obtained by replacing $\theta^*$ in~\eqref{eq:state_bellman_theta} with $\theta^{(b)}$.
This yields a greedy policy $\pi^{(b)}$. We then update
\begin{equation}
\theta^{(b+1)} = \frac{\sum_s d_{\pi^{(b)}}(s)\p{r_{R,0}(s)+\EE[X]{\pi^{(b)}(X,s)\tau_R(X,s)} } }{\sum_s d_{\pi^{(b)}}(s)\p{r_{\Delta,0}(s)+\EE[X]{\pi^{(b)}(X,s)\tau_\Delta(X,s)} } }
\end{equation}
and repeat this procedure until the transformed average outcome under $\pi^{(b)}$ is close to zero.
This can be viewed as the standard Dinkelbach update for fractional programming \citep{dinkelbach1967nonlinear},
with the inner optimization solved by the Bellman equation for the transformed outcome.

\section{Numerical Examples}
\label{sec:numerical_examples}

In this section, we study two queuing simulations motivated by concrete service systems. 
The first simulation is motivated by emergency-department fast-track triage, where the hospital decides whether to route clinically eligible patients to a fast-track queue or to the regular treatment queue. 
The second simulation is motivated by online customer support, where the platform decides whether to admit a user to the human-agent queue or route the user to an AI support channel. In the latter setting, users observe the expected wait time for human support before entering the system, so arrivals depend on the current congestion level. 

In both settings, we first compare an oracle system-aware targeting rule with a direct CADE targeting rule. These comparisons illustrate the importance of accounting for downstream congestion effects.
We then evaluate whether these system-aware rules can be learned from offline data using Algorithm~\ref{algo:augmented_cforest_general}. For the state-dependent-arrival customer-support experiment, we use the Dinkelbach-based reward-rate variant introduced in Section~\ref{subsec:reward_rate}. We refer to our implementation of Algorithm~\ref{algo:augmented_cforest_general} as state-aware CADE thresholding (SACT) in the figures, and evaluate learned policies in terms of their long-run performance.

We compare SACT with three benchmark methods. The first is a direct targeting rule that estimates the relevant state-aware CADE using a causal forest and assigns treatment whenever the estimated CADE is positive. The remaining two benchmarks are standard batch offline reinforcement-learning algorithms: discrete conservative Q-learning (CQL)~\citep{kumar2020conservative} and fitted Q-iteration (FQI)~\citep{ernst2005tree}. These reinforcement-learning baselines instantiate the collapsed-state paradigm discussed in Section~\ref{sec:comparison}: they treat the unit attributes and system state as a single Markov state, $Z_i=(X_i,S_i)$, and learn over the resulting covariate-state-action space. Thus, unlike SACT, they do not exploit the exogenous-covariate structure that separates estimation of the state-aware CADE from the lower-dimensional state-level dynamic program. Additional implementation details are provided in Appendix~\ref{subsec:implementation_details}.

\subsection{Emergency Department Fast-Track Routing}
\label{subsec:num_parallel}

Our first numerical example is motivated by fast-track systems in emergency departments. Patients arrive sequentially with observed clinical features, and triage staff must decide whether an eligible patient should be routed to a fast-track treatment stream or to the regular treatment queue. The fast-track stream provides faster service but has limited capacity. Thus, the decision is not simply whether fast-track care benefits the current patient in isolation: routing too many patients to the fast-track stream can congest that queue and reduce its value for future delay-sensitive patients. This is precisely the type of setting  where a direct targeting rule may overuse a scarce resource.

We model this setting as a parallel-queue system. Patients arrive according to an exogenous Poisson process and are routed to one of two queues: the regular queue, denoted by $j=0$, and the fast-track queue, denoted by $j=1$. Each queue $j\in\cb{0,1}$ has service rate $\mu_j$ and maximum capacity $\bk_j$. The two queues operate independently, with service proceeding whenever the corresponding queue is nonempty. Upon each arrival, we observe the patient’s covariates $X_i$ and the current queue lengths $S_i=(K_{0i},K_{1i})$. The treatment $W_i=1$ routes the patient to the fast-track queue, while $W_i=0$ routes the patient to the regular queue. Additional details on the transition dynamics of this parallel-queue system can be found in Section~\ref{subsec:parallel}. 


For each patient $i$, we observe covariates generated as $X_i \sim N(0, I_{10})$. The regular queue has capacity $\bk_0=10$, and the fast-track queue has capacity $\bk_1=3$. We set the arrival rate to $\lambda = 1$, and the service rates to $\mu_0 = 0.5$ for the regular queue and $\mu_1=1$ for the fast-track queue. Thus, the fast-track queue serves patients more quickly, but also has a smaller capacity.
Upon patient $i$'s arrival, a preliminary treatment $W_i \in \cb{0,1}$ is drawn from a $\text{Bernoulli}(0.5)$ behavior policy. At boundary states, this random draw is overridden by the feasibility rule from Section~\ref{subsec:parallel}: if one queue is full and the other is not, the patient is routed to the non-full queue; if both modeled queues are full, there is no feasible routing decision at that instant and the arrival is not included as an admitted decision epoch. The outcome $Y_i$ is generated as
\begin{align*}
Y_i = -\log(\text{waiting}_i + \text{service}_i) \cdot \mathbb{I}(X_{i,1} \le Z_{0.75}) - 3(\text{waiting}_i + \text{service}_i)^2 \cdot \mathbb{I}(X_{i,1} > Z_{0.75}) + \epsilon_i,
\end{align*}
where $\epsilon_i \sim N(0,1)$ and $Z_{0.75}$ denotes the 75th percentile of the marginal distribution of $X_{i,1}$. All patients prefer shorter waits, but the top quartile in $X_{i,1}$ is especially delay-sensitive and experiences a quadratic loss from long waiting and service times. Under this setting, the CADE of fast-track routing is positive except when the regular queue is empty and the fast-track queue is already moderately congested ($K_{0i}=0$ and $K_{1i}\ge 2$). The CADE can become very large for delay-sensitive patients, particularly when the regular queue is long.

\begin{figure}
\centering
\includegraphics[width=0.7\linewidth]{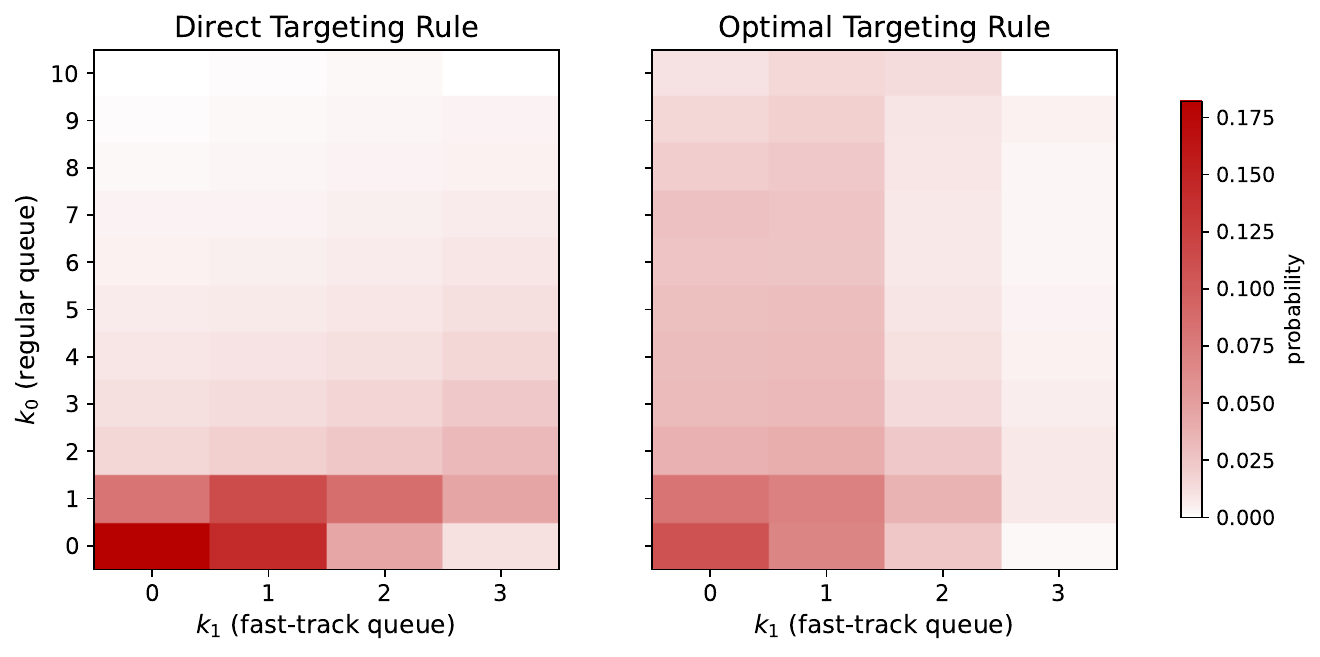}
\caption{Stationary distribution of queue lengths $(K_{0i},K_{1i})$ under the direct targeting rule and the optimal targeting rule. The intensity of the colors represents how often patients experience each combination of queue lengths, with darker colors indicating more frequent occurrences.
\ifjasa\spacingset{1}\fi
}
\label{fig:oracle_parallel}
\end{figure}

\begin{figure}
\centering
\includegraphics[width=0.8\linewidth]{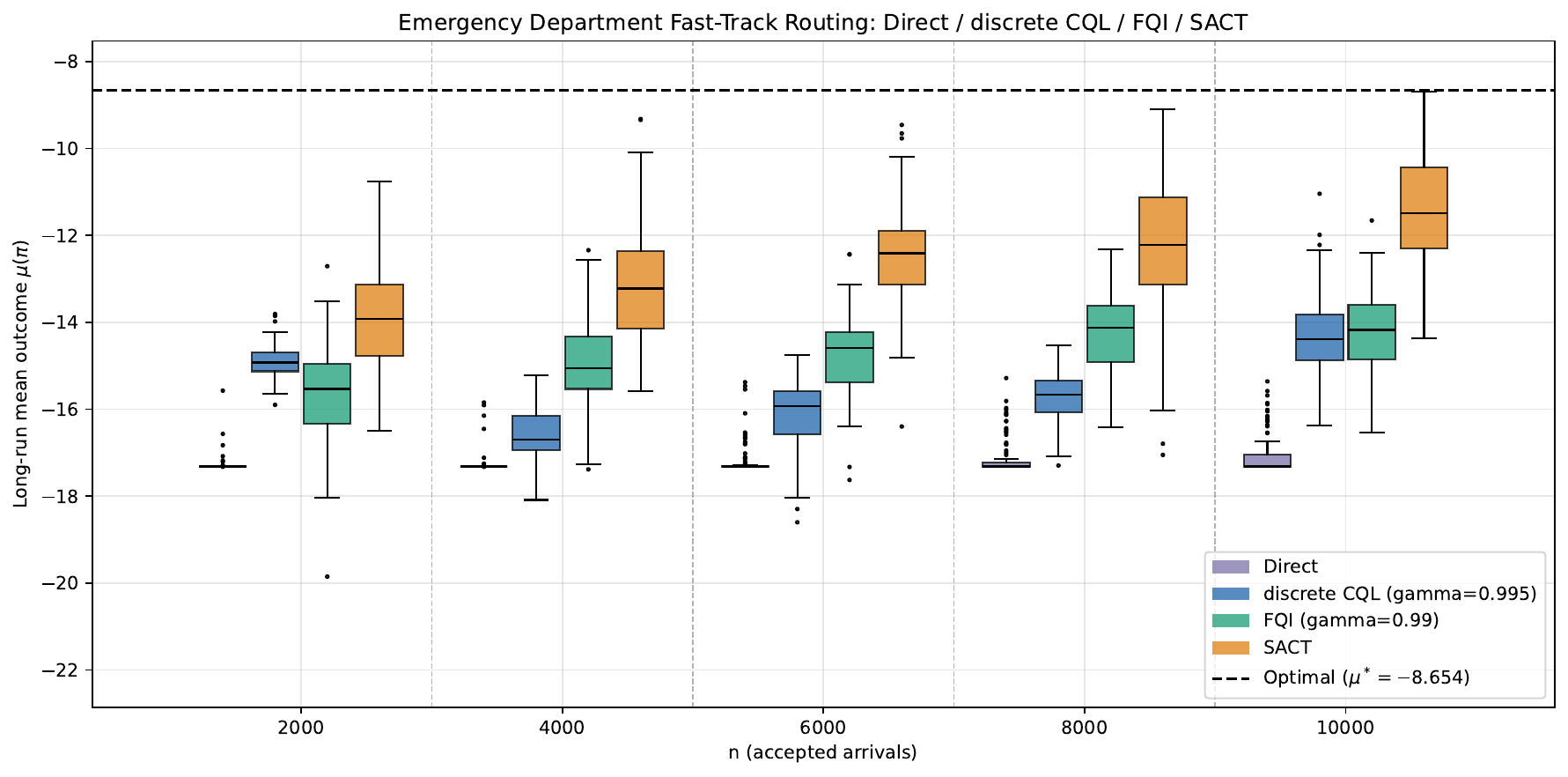}
\caption{Long-run average outcomes achieved by policies generated using Algorithm~\ref{algo:augmented_cforest_general} and the benchmark methods. The dashed horizontal line represents the optimal long-run average outcome approximated using the ground-truth model together with the policy shown in Figure~\ref{fig:oracle_parallel}.
\ifjasa\spacingset{1}\fi
}
\label{fig:policy_estimates_parallel}
\end{figure}

We compare the direct targeting rule with the oracle system-aware targeting rule. 
Figure~\ref{fig:oracle_parallel} shows that the oracle targeting rule changes the queuing system itself. Relative to direct targeting, the oracle rule shifts stationary density away from congested fast-track states such as $(k_0, 3)$ and toward less crowded states with $k_1 \in \cb{0,1,2}$. This preserves fast-track capacity so that, when delay-sensitive patients arrive, they can be routed promptly. In contrast, the direct targeting rule treats the fast-track queue as beneficial whenever the current patient's CADE is positive. Because the fast-track CADE is positive for most patient-state pairs, direct targeting tends to overutilize the fast-track queue, resulting in heavier congestion and a more imbalanced state distribution. The performance gap between direct targeting and the oracle benchmark is reported on the same long-run outcome scale in Figure~\ref{fig:policy_estimates_parallel}.

We now evaluate SACT in this emergency-department routing simulation. Figure~\ref{fig:policy_estimates_parallel} summarizes the long-run average outcomes achieved by the learned policies across $n \in \{2000,\dots,10000\}$ accepted arrivals. The direct targeting rule performs consistently poorly across all sample sizes and exhibits almost no variability across replications. This is because the estimated CADE of fast-track service is nearly always positive, leading the direct method to recover an almost always-treat policy that routes patients to the fast-track queue whenever capacity is available.

Among the benchmark reinforcement-learning methods, FQI performs better than discrete CQL and improves steadily as the sample size increases. Nevertheless, both methods remain substantially below SACT. This pattern is consistent with the fact that the reinforcement-learning baselines must learn values over the full covariate-state-action space, whereas SACT uses the structural result from Theorem~\ref{theo:policy} to reduce the dynamic component of the problem to state-specific thresholds. The performance of SACT improves steadily as $n$ increases and moves closer to the optimal policy benchmark.


\subsection{Online Customer Support with Congestion-Sensitive Arrivals}
\label{subsec:num_mnm1}

Our second numerical example is motivated by an online platform that provides customer support through two channels: a limited-capacity human-agent queue and an automated AI support channel. A user who needs help first observes an estimated wait time for a human agent and then submits a description of their problem. The platform observes request-level features, such as issue category, urgency, and predicted suitability for automated resolution. The decision is whether to admit the user to the human-agent queue ($W_i=1$) or route the user to the AI support channel ($W_i=0$), which is not capacity constrained. The reward is a customer-experience outcome observed for both channels, such as post-interaction satisfaction. 
This application differs from the emergency-department example because arrivals are congestion-sensitive. When the posted human-agent wait is long, some users defer their request or search help articles instead. The platform's routing decisions therefore affect not only the satisfaction of users who enter the system, but also the rate at which future users enter. 

We model this application using the $M_n/M/1$ queue from Section~\ref{subsec:reward_rate}. 
For each arriving user $i$, we generate covariates from a 10-dimensional normal distribution, $X_i \sim N(0, I_{10})$, where $I_{10}$ is the $10 \times 10$ identity matrix. 
Let $K_i$ denote the number of users waiting for human support when user $i$ arrives. Routing a user to the AI support channel does not increase this queue. The users' arrival rate when the queue length is $k$ is
$\lambda_k = I(k<20)\cdot 2/(k+1)^{0.1}$,
and the service rate is $\mu_k = I(k>0)$, where $k \in\{0,\dots,20\}$. The decreasing arrival rate captures the idea that users are less likely to enter the human-support flow when the visible backlog is larger. We consider a status-quo policy $\pi_0$ that admits user $i$ to the human-agent queue with probability $\pi_0(X_i,K_i)$, where
\begin{equation}
    \pi_0(x,k) = 0.6 + 0.2I(x_2>0) - 0.1 I(x_4+x_5>0),
\end{equation}
which is unknown to the policy maker.
For arriving users, the reward $R_i$ is generated as
\begin{equation}
    R_i = W_i \cdot\cb{ (7-K_i)\abs{X_{i,1}} + 3X_{i,2}} + 2\max\cb{X_{i,3},0} + \epsilon_i, \qquad \epsilon_i\sim N(0,4).
\end{equation}
The final term $2\max\cb{X_{i,3},0}$ is the baseline customer-experience reward under the AI channel. We interpret $X_{i,3}$ as measuring whether the user's issue is well suited to automated resolution. The term multiplied by $W_i$ is the incremental effect of admitting the user to the human-agent queue. The component $(7-K_i)\abs{X_{i,1}}$ makes some users especially sensitive to the human-agent wait: when the queue is short, human support can be very valuable for these users, but when $K_i>7$, the expected delay can make human support worse than immediate AI assistance. The component $3X_{i,2}$ captures issue types for which human discretion is systematically more or less valuable.

\begin{figure}[t]
\centering
\includegraphics[width=\linewidth]{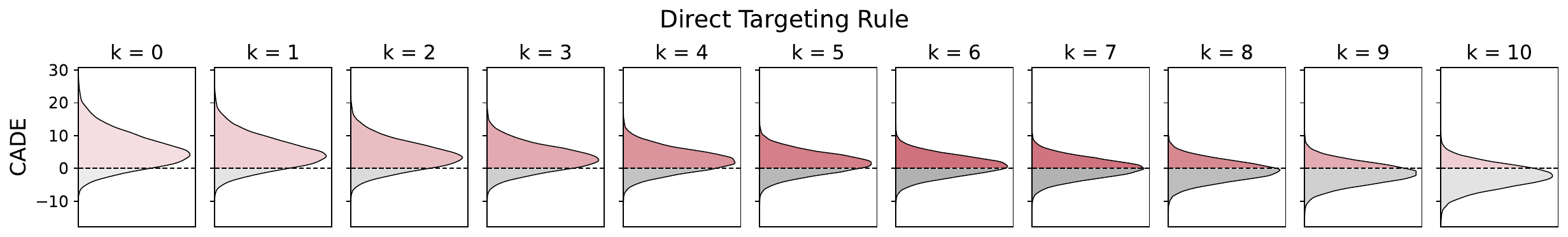}
\includegraphics[width=\linewidth]{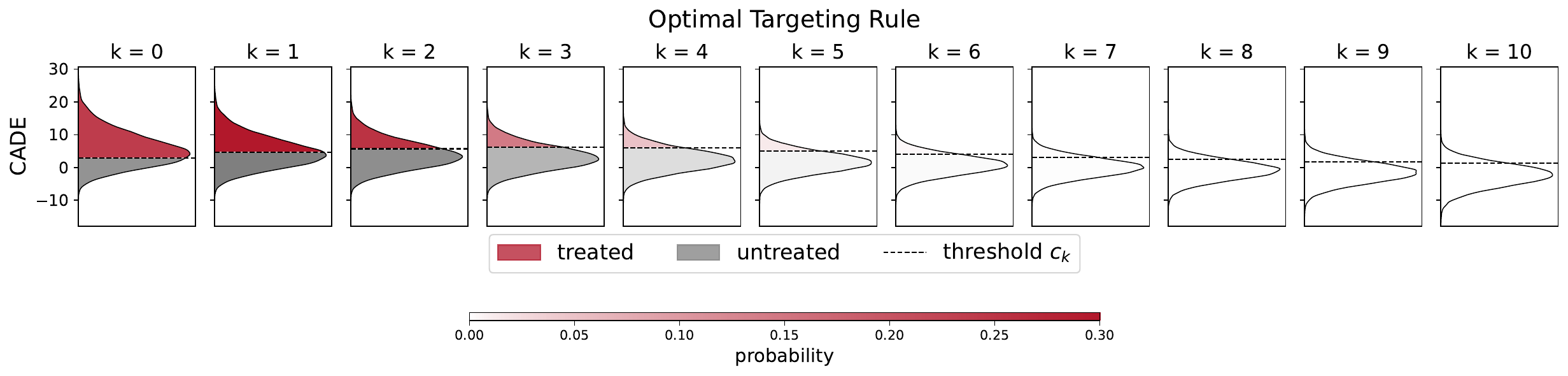}
\caption{Targeting rules for users arriving at the support system with different human-agent queue lengths ($k = 0, \dots, 10$). The full state space is $k=0,\dots,20$; states $k>10$ are omitted from the display for readability, while optimization and evaluation use all states. The curves in each panel show the distribution of the reward CADE $\tau(X_i,k)$ for users arriving when the queue length is $k$. The horizontal line represents the effective state-specific threshold for admission to the human-agent queue, which is fixed at zero under the direct targeting rule. The red area shows users admitted to human support, while the grey area shows users routed to the AI support channel. The intensity of the colors represents how often users experience each queue length, with darker colors indicating more frequent occurrences.
\ifjasa\spacingset{1}\fi
}
\label{fig:optimal_policy}
\end{figure}

A comparison between the oracle policy and the direct targeting policy is illustrated in Figure~\ref{fig:optimal_policy}. The oracle policy is more selective in admitting users to the human-agent queue, especially at smaller queue lengths where many users have positive direct effects. This may initially seem counterintuitive, since one might expect a short queue to accommodate more users. However, the objective is a long-run reward rate, and admitting marginally beneficial users can increase future waiting times and reduce the arrival rate faced by the platform. By routing users with lower CADE values to AI support, the oracle policy keeps the human queue short more often and preserves capacity for users with larger gains from timely human assistance.

Next, we examine the policy generated by SACT using the Dinkelbach-based variant introduced in Section~\ref{subsec:reward_rate}. We train each method on data observed over a time period of length $T$ and compare its long-run reward rate with the optimal policy benchmark. We compare against the same three baselines as above. To evaluate the long-run reward rate under the learned policies, we compute the conditional CADE distribution with Monte Carlo approximations and the stationary distribution analytically using the formulas provided in Section~\ref{sec:closed_form_queue} of the supplementary material. We vary $T$ from 2,000 to 10,000 and repeat the process 100 times for each horizon length considered.

\begin{figure}[t]
\centering
\includegraphics[width=0.8\linewidth]{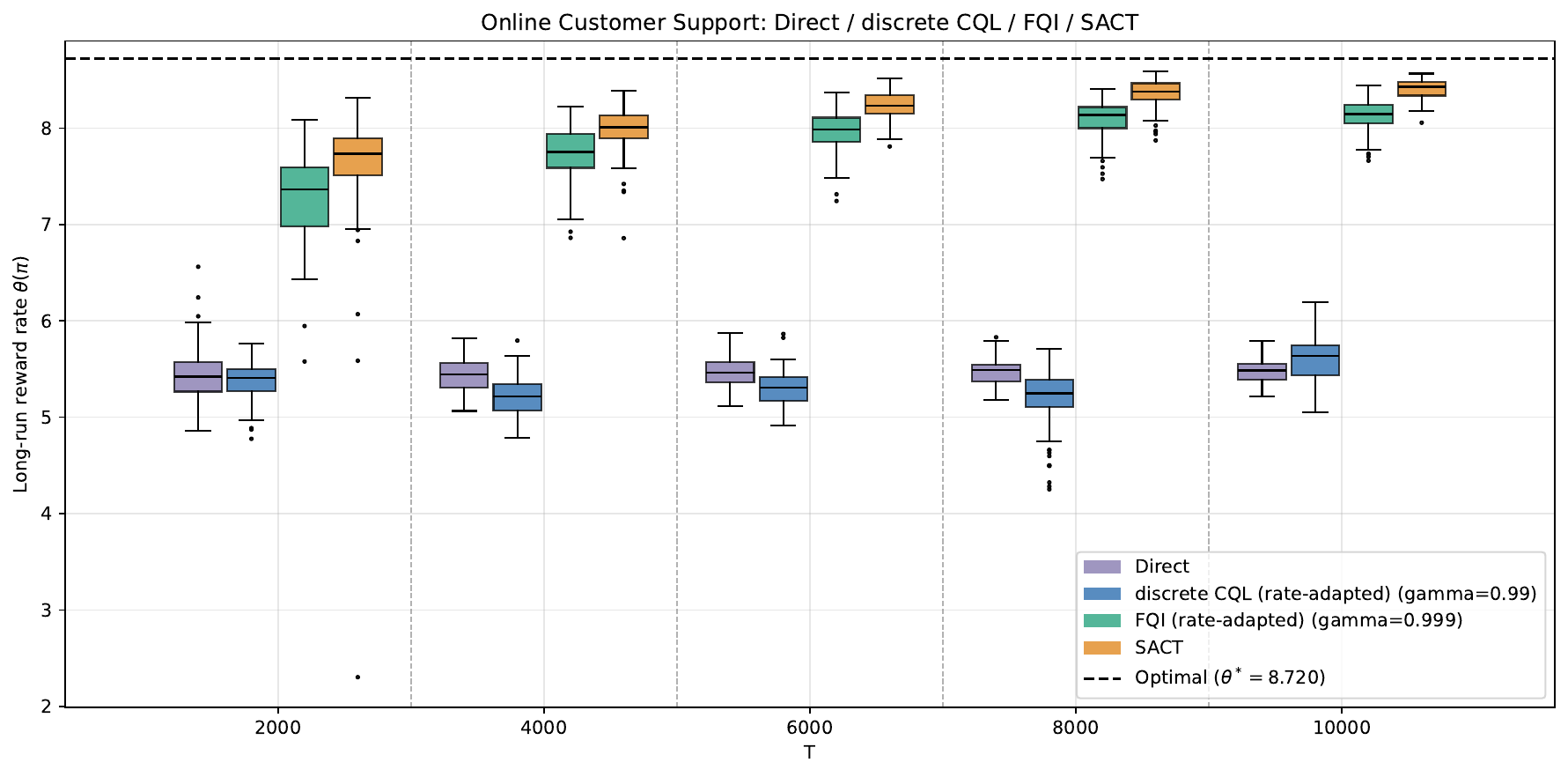}
\caption{Long-run reward rates achieved by policies generated using Algorithm~\ref{algo:augmented_cforest_general} and the benchmark methods. The dashed horizontal line represents the optimal long-run reward rate obtained using the ground-truth model together with the policy shown in Figure~\ref{fig:optimal_policy}.
\ifjasa\spacingset{1}\fi
}
\label{fig:policy_estimates}
\end{figure}

Figure~\ref{fig:policy_estimates} reports the long-run reward rates achieved by the learned policies. As the observation horizon $T$ increases, the performance of SACT steadily improves, with the resulting reward rates approaching the optimal policy benchmark indicated by the dashed line. In contrast, both the direct targeting rule and discrete CQL perform substantially worse across all horizon lengths considered. FQI performs considerably better and benefits from longer trajectories, indicating that it is able to extract useful information from the offline data. Nevertheless, its performance remains consistently below that of SACT. As in the emergency-department example, this comparison uses oracle-tuned discount factors for the reinforcement-learning baselines, whereas SACT directly targets the average-outcome objective in the exogenous-arrival setting and the reward-rate objective in the congestion-sensitive-arrival setting through state-specific thresholds.

\section{Discussion}

The existing literature on data-driven decision making is largely split into two separate strands.
One strand is focused on static, single-action settings where optimal decision can be achieved
via reasonably simple thresholding strategies \citep{athey2016recursive,manski2004statistical}
and/or empirical maximization algorithms \citep{bertsimas2020predictive,kitagawa2018should}.
The other strand considers dynamic decision making problems, for which much more complicated
reinforcement learning algorithms are typically used \citep{luckett2020estimating,robins2004optimal,sutton1999policy}.
And this dichotomy of the literature creates a challenge for practitioners. Ideally, one might
imagine building up learning algorithms step by step, where first one learns how to target under a
simplified static model and then later accommodates dynamics. But if static vs.~dynamic decision-making
problems involve completely different algorithmic solutions, then it is not clear whether it's even
possible to ``build up'' from a static targeting strategy to a dynamic one.

In this paper, we presented some results that help bridge this gap. We found that, in a
class of Markov decision processes, optimal dynamic targeting rules reduce to simple thresholding
rules just like in the static case. But, unlike in the static case, decision thresholds are
now state dependent and reflect problem dynamics---and deriving optimal thresholds requires leveraging
techniques for dynamic off-policy evaluation. Our approach thus paints a clear picture of what it
takes to move from static to dynamic targeting solutions; what insights can be kept; and what
parts of the algorithm need to be changed to accommodate dynamics. Going forward, we expect it to
be of interest to explore other decision-making problems whose difficulty lies ``between'' basic
static problems and fully general dynamic ones; and to investigate practical targeting algorithms
that can be deployed in such settings.

\ifjasa
\phantomsection\label{supplementary-material}
\bigskip
\begin{center}
{\large\bf Supplemental Material}
\end{center}
Additional application details and proofs of formal results are provided in the supplement.

\begin{center}
{\large\bf Data Availability}
\end{center}
Data sharing is not applicable to this article as no new data were created or analyzed in this study.

\fi

\ifjasa
\setlength{\bibsep}{0.2pt plus 0.3ex}

\def\spacingset#1{\renewcommand{\baselinestretch}%
{#1}\small\normalsize} \spacingset{1}
\bibliographystyle{plainnat-abbrev}
{\footnotesize
\bibliography{references}
}
\else
\bibliography{references}
\bibliographystyle{plainnat}
\fi

\newpage
\appendix
\begin{center}
\textbf{\Large Supplemental Materials} \\ 
\end{center}

\setcounter{equation}{0}
\setcounter{figure}{0}
\setcounter{table}{0}
\setcounter{page}{1}
\makeatletter
\renewcommand{\theequation}{S\arabic{equation}}
\renewcommand{\thefigure}{S\arabic{figure}}
\renewcommand{\thetable}{S\arabic{table}}
\renewcommand{\bibnumfmt}[1]{[S#1]}
\renewcommand{\citenumfont}[1]{S#1}

\section{Application: Optimal Targeting in Queuing Systems}
\label{sec:queuing}

In this section, we instantiate the general framework developed in Section~\ref{sec:general_model} in the context of queuing systems. We begin with a simple admission problem in an $M$/$M$/$1$ queue with finite capacity $\bar{k}$,\footnote{\ifjasa\spacingset{1}\footnotesize\fi
Alternatively, the system can be viewed as a queue without capacity constraints under a policy that assigns patients to control whenever the queue length reaches $\bar{k}$.} where the policy determines which units are admitted into the queue. 
We then discuss how the same approach extends to parallel queues, where the policy routes each arriving unit to one of several service tracks. These examples illustrate how the state-aware CADE
thresholding rule accounts for the downstream congestion effects induced by current admission decisions.

\subsection{Targeting in a Simple Queuing Model}
\label{subsec:mm1}

Consider an $M/M/1$ queue in which patients arrive at rate $\lambda$ and are served at rate $\mu$, where both rates are bounded and bounded away from zero \citep{ross2014introduction}. We observe a sequence of $n$ patient arrivals, indexed by $i=1,\dots,n$, under a status quo behavior policy $\pi_0$, which could be potentially unobserved.
When patient $i$ arrives, we observe their covariates $X_i\in\mathcal{X}$ and the current queue length $K_i=0,1,\dots,\bk$. 
The system then assigns treatment $W_i\in\{0,1\}$, where $W_i=1$ corresponds to admitting the patient into the
queue and $W_i=0$ corresponds to assigning the patient to the outside option. 
We record the outcome $Y_i$, which may depend on the patient's covariates, the treatment assignment, and the queue length at the time of arrival.\footnote{\ifjasa\spacingset{1}\footnotesize\fi
Although $Y_i$ may depend on the realized waiting time, under an $M/M/1$ queue with a state-independent service rate $\mu$, the distribution of the realized waiting time for an admitted patient is determined by the queue length at the time of arrival and the admission decision. Thus, the conditional law of $Y_i$ can still be written as a function of $(X_i,K_i,W_i)$, as required by Assumption~\ref{assu:MDP}.} As in Assumption~\ref{assu:stationary}, we assume that $Y_i$ is bounded almost surely.

In such a queuing system, the queue length $K_i$ plays the role of the system state. Conditional on the current queue length and treatment assignment, the next queue length is determined by whether the current patient is admitted and by the number of service completions before the next arrival. Thus, when the process is observed at arrival epochs, $\{K_i\}$ forms a finite-state Markov chain whose transition kernel depends on the current queue length and action. In this way, the simple queuing model
satisfies the exogenous-covariate Markov structure in Assumption~\ref{assu:MDP}, with $S_i=K_i$.

In this setting, Theorem~\ref{theo:policy} implies that an optimal admission policy takes the form $\pi^*(x,k)=I(\tau(x,k)>c_k )$, where the threshold $c_k$ depends on the current queue length. 
Admitting the current patient affects not only that patient's outcome, but also the queue-length process faced by future arrivals until subsequent service completions relieve the
system, and the threshold $c_k$ internalizes this downstream congestion effect. 
Thus, the state-aware rule may reject a patient with positive direct benefit, not because the treatment is individually harmful, but because the capacity used by that admission has greater expected value for future arrivals.
Algorithm~\ref{algo:augmented_cforest_general} can then be applied directly with $S_i=K_i$. In particular, we estimate the CADE $\tau(x,k)$, the baseline reward $r_0(k)$, and the queue-length transition kernels $P_{K,w}(\cdot\mid k)$, and then run
relative value iteration over the finite queue-length state space. 

When the maximum queue length $\bk$ is large, modeling the thresholds $c_k$ as functions of $k$ can sometimes lead to improved finite-sample performance. This parallels common practice in policy learning for MDPs, where treatment probabilities $\pi(x, k)$ are often modeled as functions of both $x$ and $k$ \citep{sutton2018reinforcement}. As in those settings, parametrization introduces additional approximation error depending on how well the chosen functional form captures the true thresholds. However, since our policy space is much simpler than that of general MDPs, the quality of approximation is often better in practice.

\subsection{Optimal Routing with Parallel Queues}
\label{subsec:parallel}

In the basic setting considered above, the policy determines whether to admit an individual into a single queue. In many practical systems, however, individuals can be routed to parallel queues with different service speeds, and the decision lies in routing each arriving individual to one of the available service tracks. For example, in emergency departments, triage nurses need to decide whether to route a patient to a fast-track urgent care line or a regular treatment line; similarly, in online delivery platforms, the system needs to decide whether to assign a customer to expedited delivery or assign them to standard delivery.

We show how our framework can be extended to accommodate such parallel queuing systems by augmenting the state space, where routing decisions influence not only individual outcomes but also system dynamics in both of the queues. In such a parallel queuing system, individuals arrive according to a Poisson process with rate $\lambda$ and are routed to one of two queues, where each queue $j\in \cb{0,1}$ has its own service rate $\mu_j$ and maximum capacity $\bk_j$. The two queues operate independently, and service proceeds only if the respective queue is non-empty. 

Upon each admissible arrival, we observe the individual's covariate $X_i$ as well as the current queue lengths $(K_{0i},K_{1i})$, and assign a treatment $W_i\sim\pi(X_i,K_{0i},K_{1i})$ indicating whether the individual is routed to queue 0 or 1. 
When both queues have available capacity, the policy chooses where to route the individual.
If one queue is full and the other is not, the individual is routed to the non-full queue; if both queues are full, potential arrivals are blocked by capacity and are not
included in the decision process. Thus, the learned threshold rule is applied only at states where both routing actions are feasible, while boundary decisions are determined by the capacity constraints.
After assignment, we record the outcome $Y_i$, which could depend on the individual's covariates and the realized wait and service times in the assigned queue. As before, we assume $Y_i$ is bounded almost surely.

To verify that our framework applies to this setting, we note that this is a discrete-time Poisson point process observed at admissible arrivals. Under a fixed routing policy $W_i\sim \pi$, the evolution of the queue lengths $(K_{0i},K_{1i})$ follows a time-homogeneous Markov decision process such that
\begin{equation}
\begin{split}
    K_{0,i+1} &= K_{0i} + (1-W_i) \cdot I(K_{0i}<\bk_0) + W_i \cdot I(K_{1i}=\bk_1) -N_{\text{served},0,i},\\
    K_{1,i+1} &= K_{1i} + W_i \cdot I(K_{1i}<\bk_1) + (1-W_i) \cdot I(K_{0i}=\bk_0) -N_{\text{served},1,i},
\end{split}    
\end{equation}
where $N_{\text{served},0,i}$ and $N_{\text{served},1,i}$ denote the actual numbers of individuals served between the $i$th and the $i+1$th admissible arrivals in queue 0 and queue 1. We refer readers to Section \ref{sec:closed_form_queue} of the supplementary material for details on the state dynamics of this system.

Since arrivals and services follow Poisson processes, this embedded process recorded at arrivals satisfies Assumptions \ref{assu:MDP}-\ref{assu:stationary}. 
Algorithm \ref{algo:augmented_cforest_general} can then be applied directly with $S_i=(K_{0i},K_{1i})$. 
In Section \ref{subsec:num_parallel}, we demonstrate through simulations that the extended algorithm learns a system-aware targeting policy that always outperforms the direct targeting benchmark and converges toward the optimal policy as the sample size increases.


\begin{rema}
By augmenting the state space, our framework can also accommodate dynamic resource allocation problems. Upon each arrival, the system observes both the current resource level and the covariates of the incoming individual including their resource demand, and must decide whether to approve the request. This setup fits naturally into our framework by defining the state as the pair of the current resource level and the individual’s demand. This formulation also captures scenarios with multiple types of operations that require different service times (e.g., major versus minor medical procedures, long-distance versus short-distance deliveries), where each type corresponds to a different resource demand level of the service.
\end{rema}

\section{Closed Forms of Transition Kernels and Stationary Distributions}
\label{sec:closed_form_queue}

In this section, we derive the closed forms of transition kernels and stationary distributions for the queuing processes considered in the paper. We start with the transition kernel and stationary distribution of the embedded point process recorded at both arrivals and services corresponding to the $M_n/M/1$ system discussed in Section \ref{subsec:reward_rate}, with the standard $M/M/1$ model in Appendix~\ref{subsec:mm1} as a special case. We then show that it is possible to find the stationary distribution of the embedded point process recorded only at arrivals using the stationary distribution of the embedded point process recorded at both arrivals and services. Finally, we derive the closed forms of transition kernels of the parallel queuing system discussed in Section \ref{subsec:parallel}, which provides foundation for approximating its stationary distribution computationally.

\subsection{\texorpdfstring{Transition Kernel and Stationary Distribution of an $M_n/M/1$ system}{Transition Kernel and Stationary Distribution of an Mn/M/1 system}}

For the embedded point process recorded at both arrivals and services, the state $S_i$ contains both the event type $A_i$ and the queue length $K_i$. Recall that $A_i=1$ represents arrivals (i.e., a unit arriving at the system) while $A_i=0$ represents services (i.e., a unit leaving the system). Let $\bpi(k) = \EE{\pi(X_i,K_i)\cond K_i=k, A_i=1}$ be the probability of assigning a unit to the queue given current queue length $k$. 

First, we note that $P^\pi_{S}(s'\cond s)$, the system's transition kernel under policy $\pi$, can be written as
\begin{equation}
\begin{split}
&P^\pi_{S}(s'\cond s)\\
&\quad= P^\pi_{S}(a',k'\cond a,k)\\
&\quad= \bpi(k)\PP{A_{i+1}=a',K_{i+1}=k'\cond A_i=a,K_i=k,W_i=1}\\
&\quad\quad + (1-\bpi(k))\PP{A_{i+1}=a',K_{i+1}=k'\cond A_i=a,K_i=k,W_i=0}\\
&\quad= \bpi(k)\PP{K_{i+1}=k'\cond A_i=a,K_i=k,W_i=1}\PP{A_{i+1}=a'\cond K_{i+1}=k'}\\
&\quad\quad + (1-\bpi(k))\PP{K_{i+1}=k'\cond A_i=a,K_i=k,W_i=0}\PP{A_{i+1}=a'\cond K_{i+1}=k'},
\end{split}
\end{equation}
where the last equality holds because the probability that the next event is an arrival or not depends only on the current queue length. Furthermore, 
\begin{equation}
\PP{A_{i}=a\cond K_{i}=k} = \begin{cases}
    &{\lambda_k}I\p{k<\bk}/(\lambda_kI\p{k<\bk}+\mu I\p{k>0})\qquad \text{ if }a=1\\
    &{\mu}I\p{k>0}/(\lambda_kI\p{k<\bk}+\mu I\p{k>0})\qquad \text{ if }a=0.
\end{cases}
\label{eq:arrival_transition}
\end{equation}

To derive $\PP{K_{i+1}=k'\cond A_i=a,K_i=k,W_i=w}$, observe that the queue length at the next event is deterministic given the current queue length, the event type, and the treatment assignment. In particular, the probability above is nonzero (and equal to one) only in the following cases:
\begin{itemize}
    \item If $a=0$ and $k'=k-1$, then the most recent event is a service completion, which decreases the queue length by one.
    \item If $a=1$, $w=0$ and $k'=k$, then the most recent event is an arrival, but the unit is not assigned to the queue. Thus, the queue length remains unchanged.
    \item If $a=1$, $w=1$ and $k'=k+1$, then the most recent event is an arrival and the unit is assigned to the queue. Thus, the queue length increases by one.
\end{itemize}
Putting everything together, 
\begin{equation}
P^\pi_{S}(a',k'\cond a,k) = \begin{cases}
    &\bpi(k)\PP{A_{i}=a'\cond K_{i}=k+1}\qquad \text{ if } a=1, k'=k+1\\
    &(1-\bpi(k))\PP{A_{i}=a'\cond K_{i}=k}\qquad \text{ if }a=1, k'=k\\
    &\PP{A_{i}=a'\cond K_{i}=k-1}\qquad \text{ if } a=0, k'=k-1\\
    &0, \qquad \text{ otherwise},
\end{cases}
\end{equation}
where the form of $\PP{A_{i}=a\cond K_{i}=k}$ is given in \eqref{eq:arrival_transition}.

To compute the stationary distribution $d_{\pi}(a,k)$,
notice that the next event type is sampled independently of the past given the queue length. This allows us to first focus on the dynamics of the queue lengths only and derive $d^\dag(k)$. $d(a,k)$ could then be computed as 
\begin{equation}
    d(a,k)=d^\dag(k)\cdot \PP{A_{i}=a\cond K_{i}=k}.
\end{equation}
Since the number of states is finite, solving the corresponding system of equations to obtain the stationary distribution is straightforward, with
\begin{equation}
    d^\dag(k) = \frac{r_k}{\sum_{m=0}^{\bk} r_k}, \qquad k=0,\dots,\bk,
\end{equation}
and
\begin{equation}
\begin{split}
& r_0 = 1,\\
& r_1 = \frac{(\lambda_1+\mu)\bpi_0}{\mu}, \\
& r_k = \frac{(\lambda_k+\mu)\bpi_0}{\mu} \prod_{m=1}^{k-1} \frac{\lambda_m\bpi_m}{\mu} , \qquad k=2,\dots,\bk-1,\\
& r_{\bk} = \bpi_0 \prod_{m=1}^{\bk-1} \frac{\lambda_m\bpi_m}{\mu}.
\end{split}
\label{eq:stationary_r}
\end{equation}

\subsection{Stationary Distribution of Embedded Point Process Recorded at Arrivals}

To find the stationary distribution of the embedded point process recorded only at arrivals, we note that this process further conditions on the event being an arrival. Thus, its stationary distribution could be expressed and computed as
\begin{equation}
\begin{split}
\PP[\pi]{K=k\cond A=1} 
&= \frac{\PP[\pi]{K=k,A=1}}{\sum_{k'=0}^{\bk}\PP[\pi]{K=k',A=1}}\\
&= \frac{d(1,k)}{\sum_{k'=0}^{\bk}d(1,k')}.
\end{split}
\end{equation}

\subsection{Parallel Queue with Routing}



Let the state be $(A,K_0,K_1)$, where $A=0$ and $A=1$ represent service at queues 0 and 1, respectively, and $A=2$ represents an arrival. For simplicity, we further denote 
$$f_{A\cond K_0,K_1}(a\cond k_0,k_1) := \PP{A_{i}=a\cond K_{0i}=k_0, K_{1i}=k_1}.$$
It is possible to write down the transition kernel for this setting:
\begin{equation*}
P_{S}(a',k_0',k_1' \cond 2,k_0,k_1,w) = \begin{cases}
    &f_{A\cond K_0,K_1}(a'\cond k_0',k_1')\qquad \text{ if } k_0'=k_0+1,k_1'=k_1,w=0\\
    &f_{A\cond K_0,K_1}(a'\cond k_0',k_1')\qquad \text{ if } k_0'=k_0,k_1'=k_1+1,w=1\\
    &0, \qquad \text{ otherwise},
\end{cases}
\end{equation*}
\begin{equation*}
P_{S}(a',k_0',k_1' \cond 0,k_0,k_1,w) = \begin{cases}
    &f_{A\cond K_0,K_1}(a'\cond k_0',k_1')\qquad \text{ if } k_0'=k_0-1,k_1'=k_1,k_0>0\\
    &0, \qquad \text{ otherwise},
\end{cases}
\end{equation*}
and
\begin{equation*}
P_{S}(a',k_0',k_1' \cond 1,k_0,k_1,w) = \begin{cases}
    &f_{A\cond K_0,K_1}(a'\cond k_0',k_1')\qquad \text{ if } k_0'=k_0,k_1'=k_1-1,k_1>0\\
    &0, \qquad \text{ otherwise},
\end{cases}
\end{equation*}
where
\begin{equation*}
f_{A\cond K_0,K_1}(a\cond k_0,k_1) = \begin{cases}
    &\widetilde\lambda / (\widetilde\lambda + \mu_0 I\p{k_0 > 0} + \mu_1 I\p{k_1 > 0})\qquad \text{ if }a=2\\
    &\mu_0 I\p{k_0 > 0} / (\widetilde\lambda + \mu_0 I\p{k_0 > 0} + \mu_1 I\p{k_1 > 0})\qquad \text{ if }a=0\\
    &\mu_1 I\p{k_1 > 0} / (\widetilde\lambda + \mu_0 I\p{k_0 > 0} + \mu_1 I\p{k_1 > 0})\qquad \text{ if }a=1,
\end{cases}
\end{equation*}
and $\widetilde\lambda=\lambda \cdot I\p{k_0 < \bk_0 \text{ or } k_1 < \bk_1}$. The system's transition kernel under policy $\pi$ can be thus expressed as follows:
\begin{equation*}
P^\pi_{S}(a',k_0',k_1'\cond a,k_0,k_1) = \begin{cases}
    &\bpi(k_0,k_1)f_A(a' \mid k_0, k_1 + 1) \qquad \text{ if }a=2, k_0'=k_0, k_1'=k_1+1\\
    &(1-\bpi(k_0,k_1))f_A(a' \mid k_0 + 1, k_1)\qquad \text{ if }a=2, k_0'=k_0+1, k_1'=k_1\\
    &f_A(a' \mid k_0, k_1 - 1)\qquad \text{ if }a=1,k_0'=k_0,k_1'=k_1-1,k_1>0\\
    &f_A(a' \mid k_0 - 1, k_1)\qquad \text{ if }a=0,k_0'=k_0-1,k_1'=k_1,k_0>0\\
    &0, \qquad \text{ otherwise}.
\end{cases}
\end{equation*}

\section{Simulation Details}
\label{subsec:implementation_details}

This section provides implementation details for the simulation studies in Section~\ref{sec:numerical_examples}. 

\subsection{Oracle Benchmarks}
For the oracle comparisons in Figures~\ref{fig:oracle_parallel} and~\ref{fig:optimal_policy}, the direct targeting rule thresholds the true state-aware CADE at zero. The oracle system-aware rule is obtained by searching over threshold policies and evaluating each candidate under the ground-truth queuing model.

In the emergency-department experiment, we search over state-specific threshold policies using the COBYLA algorithm~\citep{Powell1994,powell1998direct}, as implemented in \texttt{scipy.optimize.minimize} from the Python \texttt{SciPy} package~\citep{virtanen2020scipy}. The long-run average outcome is computed under the ground-truth data-generating process, and the stationary distribution is obtained from the transition matrix derived in Section~\ref{sec:closed_form_queue}.

In the online customer-support experiment, we again search over threshold policies, but impose an additional monotonicity constraint on the induced marginal admission probability. Specifically, writing
$\bar\pi(k) = \PP{\pi(X_i,k)=1}$,
we require $\bar\pi(k_1)\le \bar\pi(k_2)$ whenever $k_1\ge k_2$. This constraint is not required by Theorem~\ref{theo:policy_consistency}; it is used only to stabilize the finite-dimensional oracle search and to rule out highly oscillatory threshold vectors. The resulting policy should therefore be interpreted as an approximate oracle benchmark within this constrained threshold class. The constrained optimization is carried out using COBYLA, a derivative-free method for nonlinear optimization with inequality constraints~\citep{Powell1994,powell1998direct}, as implemented in \texttt{scipy.optimize.minimize}~\citep{virtanen2020scipy}.

\subsection{Offline Training and Policy Evaluation}
For the learned-policy comparisons, all methods are trained on the same offline trajectories within each simulation replication. The performance boxplots in Figures~\ref{fig:policy_estimates_parallel} and~\ref{fig:policy_estimates} are based on 100 independent simulation replications for each sample size or observation horizon. Learned policies are evaluated under the ground-truth data-generating process to obtain their long-run average outcome or reward rate. This ground-truth model is used only for oracle benchmarking and final policy evaluation; it is not used to train SACT, the direct CADE baseline, CQL, or FQI.

\subsubsection{SACT and the Direct CADE Baseline}
All causal forests used in the direct baseline and in SACT are fit with 500 trees. The direct baseline estimates the state-aware CADE and assigns treatment whenever the estimated CADE is positive. SACT uses the same CADE-estimation specification, but combines the estimated CADEs with state-specific continuation-value thresholds obtained from the state-level dynamic program.

For SACT, we use two-fold cross-fitting for the CADE and nuisance estimates entering the empirical Bellman operator. Because the observations come from dependent trajectories, we do not split adjacent observations by independently shuffling individual arrivals. Instead, folds are constructed from trajectory blocks. In the reported simulations, natural empty-system states are used to form regenerative blocks when available, and the resulting blocks are randomly assigned to the two folds.

The state-level transition kernels used by SACT are estimated from the offline data by empirical conditional transition frequencies. They are not replaced by ground-truth transition matrices during training. In the emergency-department experiment, the state is the pair of queue lengths $(K_{0i},K_{1i})$, where $K_{0i}$ is the regular-queue length and $K_{1i}$ is the fast-track-queue length. In the customer-support experiment, the state is the human-agent queue length $K_i$. Relative value iteration is initialized at $V^{(0)}\equiv 0$ and recentered at an empty-system anchor state: $(K_0,K_1)=(0,0)$ in the emergency-department experiment and $K=0$ in the customer-support experiment. We iterate until the sup-norm change in the recentered value function is below $10^{-6}$ or until 2,000 iterations have been reached.

\subsubsection{Reinforcement-Learning Baselines}
For the reinforcement-learning baselines, FQI is implemented using the \texttt{mushroom\_rl} implementation with 25 fitted-Q iterations and an \texttt{ExtraTreesRegressor} backend with 120 trees. Discrete CQL is implemented using \texttt{d3rlpy}, with a two-layer neural encoder with 64 hidden units per layer, 120 training epochs, learning rate $3\times 10^{-4}$, batch size 128, and conservative penalty $\alpha=0.2$.

Both CQL and FQI are trained on arrival-level decision tuples. In the emergency-department experiment, the state features are $(X_i,K_{0i},K_{1i})$. In the customer-support experiment, the state features are $(X_i,K_i)$. The next state is the state observed at the next arrival decision epoch. Service completions between arrivals determine the next queue length and, in the customer-support experiment, the elapsed calendar time, but service completions themselves are not treated as decision epochs for the offline learners.

In the emergency-department experiment, the arrival rate is constant, so no reward-rate correction is needed. The reinforcement-learning baselines are therefore trained directly on the realized outcome $Y_i$. Boundary constraints are imposed for all methods in this setting: if the fast-track queue is full, the policy routes the patient to the regular queue; if the regular queue is full, the policy routes the patient to the fast-track queue; and if both modeled queues are full, the arriving patient is not included as an admitted decision epoch. The same feasibility override is applied during data generation, policy learning, and policy evaluation.

For CQL and FQI, we report oracle-tuned discount factors. Specifically, we search over
$\gamma \in \{0.95,0.98,0.99,0.995,0.999\}$
and, for each method and environment, report the value of $\gamma$ with the best average ground-truth long-run performance across sample sizes and replications. This tuning selects $\gamma=0.99$ for customer-support CQL, $\gamma=0.999$ for customer-support FQI, $\gamma=0.995$ for emergency-department CQL, and $\gamma=0.99$ for emergency-department FQI, matching the legends in Figures~\ref{fig:policy_estimates_parallel},~\ref{fig:policy_estimates},~\ref{fig:runtime_parallel_queues}, and~\ref{fig:runtime_single_queue}. Because this tuning uses ground-truth long-run performance, it is favorable to the reinforcement-learning baselines and would not be available in a real offline-learning application.

\subsubsection{Reward-Rate Implementation for the Customer-Support Experiment}
For the customer-support experiment, the objective is the long-run reward rate
$\theta(\pi)=\EE[\pi]{R_i}/\EE[\pi]{\Delta_i}$,
where $\Delta_i$ denotes the elapsed calendar time between consecutive user-arrival decision epochs in the arrival-embedded implementation. We therefore use the Dinkelbach version of SACT described in Section~\ref{subsec:reward_rate}. In the implementation, the reward CADE $\tau_R$, the elapsed-time CADE $\tau_\Delta$, and the state transition kernels are estimated once from the offline data. The Dinkelbach loop then updates the transformed outcome $R_i-\theta\Delta_i$ without refitting the causal forests. We initialize $\theta$ at the empirical reward rate $\sum_i R_i/\sum_i \Delta_i$ computed on the training portion of the trajectory, and stop when the estimated transformed average reward is below $10^{-5}$ in absolute value or after 50 Dinkelbach updates.

We adapt CQL and FQI to the same reward-rate objective by training them on transformed rewards
$\widetilde R_i(\rho) = R_i - \rho \Delta_i$
over a grid of candidate values of $\rho$. Let
$\hat\theta_{\mathrm{emp}}=\sum_i R_i/\sum_i \Delta_i$
denote the empirical reward rate computed on the training portion of the data. We use the five-point grid
$\rho \in \{0,0.5\hat\theta_{\mathrm{emp}},\hat\theta_{\mathrm{emp}},
1.5\hat\theta_{\mathrm{emp}},2\hat\theta_{\mathrm{emp}}\}$.
We split the observed arrivals chronologically into a 70\% training portion and a 30\% validation portion. For each value of $\rho$, we train the off-the-shelf CQL or FQI learner on the transformed reward and select the policy with the largest validation estimate of $\theta(\pi)$. The validation estimate is computed by importance weighting using an estimated behavior policy $\hat\pi_0(X_i,K_i)$, clipped to $[0.05,0.95]$. In the customer-support experiment, this estimated behavior model is allowed to depend on both user covariates and human-agent queue length, although the true status-quo policy specified in Section~\ref{subsec:num_mnm1} depends only on user covariates.

\subsubsection{Runtime Measurement}
Runtime is measured as training time only and excludes data simulation, computation of the optimal policy benchmark, and evaluation. For reward-rate versions of CQL and FQI, the reported runtime includes training over the candidate $\rho$ grid and selecting the final policy using the validation estimate described above.

\begin{figure}[t]
	\centering
	\includegraphics[width=\linewidth]{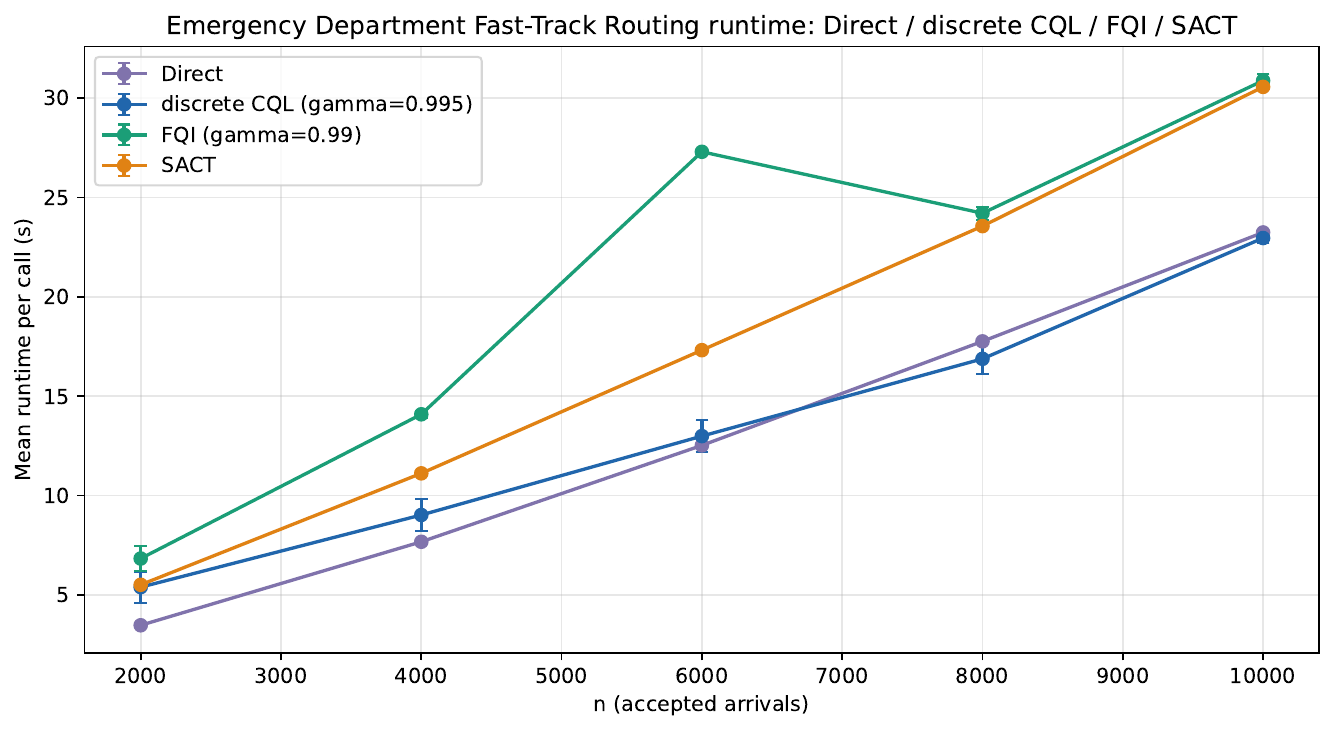}
	\caption{Runtime in the emergency-department fast-track routing experiment. The plot reports mean training time per call, excluding data simulation and evaluation. The gap between FQI and the other methods is much smaller than in the customer-support experiment because no reward-rate grid search is needed.}
	\label{fig:runtime_parallel_queues}
\end{figure}

Figure~\ref{fig:runtime_parallel_queues} reports runtime results for the emergency-department fast-track routing experiment. The methods have broadly comparable runtimes. FQI and SACT are somewhat more computationally expensive than the direct approach and discrete CQL, but they also achieve stronger empirical performance in Figure~\ref{fig:policy_estimates_parallel}.

Reported runtimes are single-CPU SLURM wall-clock times. Each configuration was submitted as a separate SLURM job, and the 100 replications within a configuration were executed sequentially on the same compute node. Because different configurations may run on different nodes, a small amount of hardware-induced variability remains in the reported runtimes. This variability is most visible for FQI, whose runtime is dominated by repeatedly fitting large tree ensembles. As a result, the FQI curve is not perfectly monotone in sample size; for example, the runtime at $n=6000$ is slightly larger than at $n=8000$. We view this as measurement noise arising from differences across compute nodes rather than a meaningful algorithmic phenomenon. Overall, runtime increases with sample size for all methods.

\begin{figure}[t]
	\centering
	\includegraphics[width=\linewidth]{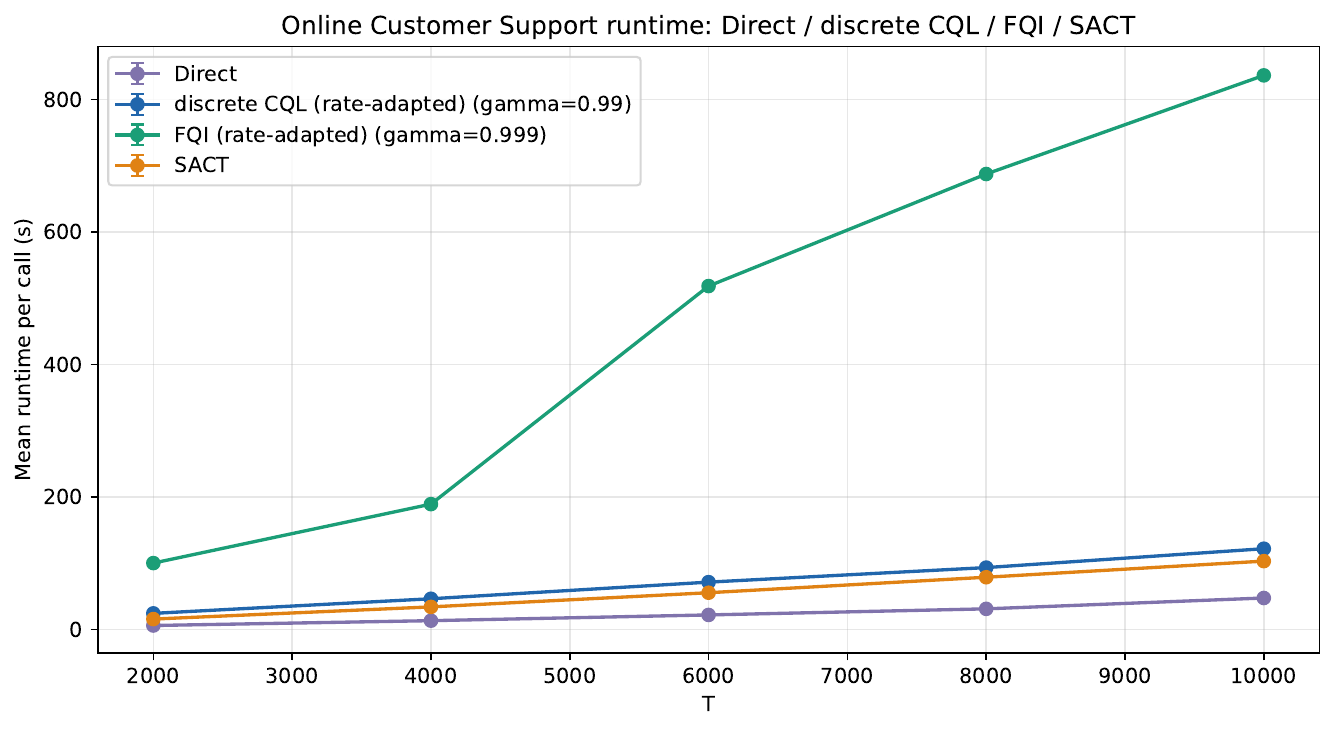}
	\caption{Runtime in the online customer-support experiment. The plot reports mean training time per call, excluding data simulation and evaluation. FQI is substantially slower than the other methods because the reward-rate adaptation requires retraining the full fitted-Q procedure over a grid of transformed rewards.}
	\label{fig:runtime_single_queue}
\end{figure}

Figure~\ref{fig:runtime_single_queue} reports runtime results for the online customer-support experiment. All methods become more expensive as the observation horizon increases, but the increase is much sharper for FQI. This is mainly due to the reward-rate adaptation. For each candidate value of $\rho$, FQI runs a full sequence of fitted-Q iterations, and each iteration refits a 120-tree ensemble to Bellman targets based on the transformed reward $R_i-\rho\Delta_i$. Since the elapsed time $\Delta_i$ is state-dependent in the customer-support system, changing $\rho$ can substantially change the scale and shape of the regression target. Thus, the five-point $\rho$ sweep is especially costly for the tree-based FQI implementation. CQL is also retrained over the same grid, but its neural-network architecture and number of gradient steps are fixed, so the cost of each run is less sensitive to the transformed reward scale. SACT also solves the reward-rate problem iteratively, but its nuisance components are estimated once and reused across Dinkelbach updates, which keeps the runtime substantially below FQI.

\section{Proof of Theorems, Propositions, and Corollaries}

\subsection{Proof of Theorem~\ref{theo:policy}}

From Lemma~\ref{lemm:bellman_equation_state}, any deterministic policy $\pi^*$ that satisfies
\begin{equation}
\pi^*(x,s) \in \argmax_{w \in \cb{0,1}} \cb{{\eta}_w(X,s) +\sum_{s'\in\mathcal S} P_{S,w}(s'|s) V^*(s')}
\end{equation}
for all $(x,s)\in\mathcal X\times\mathcal S$ is optimal.
Note that for each fixed $(x,s)$, treatment is strictly preferred to control if and only if
\begin{equation}
\tau(x,s) + \sum_{s'\in\mathcal S}
\p{P_{S,1}(s'\cond s)-P_{S,0}(s'\cond s)}V^*(s') >0.
\end{equation}
Equivalently, treatment is strictly preferred whenever $\tau(x,s)>c_s$, where 
\begin{equation}
c_s = - \sum_{s'\in\mathcal{S}} \p{P_{S,1}(s'\mid s)-P_{S,0}(s'\mid s)}V^*(s')
\end{equation}
At equality, the two actions have the same Bellman value, so choosing control still attains the maximum. Thus, the strict-threshold rule $\pi^*(x,s)=I\p{\tau(x,s)>c_s}$ attains the pointwise maximum and is optimal.

\subsection{Proof of Theorem~\ref{theo:policy_consistency}}

Recall that the oracle policy thresholds the difference-of-Qs at zero:
\begin{equation}
\begin{split}
\pi^*(x,s) := &I\p{\delta^*(x,s)\ge 0 },\\
\delta^*(x,s):=&Q_{\pi^*}(x,s,1) - Q_{\pi^*}(x,s,0) \\
=&\tau(x,s)
+
\sum_{s'\in\mathcal S}\{P_{S,1}(s'\cond s)-P_{S,0}(s'\cond s)\}V^*(s'),
\end{split}
\end{equation}
with the differential Q-function under policy $\pi$ defined as
\begin{equation}
Q_{\pi}(x,s,w) = \lim_{n\to\infty} \EE[\pi]{\sum_{i=1}^{n} \p{Y_i-\mu(\pi)}\cond X_1=x,S_1=s,W_1=w}
\label{eq:Q_xsw_defi}
\end{equation}

The estimated policy can be written as
\begin{equation}
\hpi(x,s) := I(\hdelta(x,s)\ge 0)
\end{equation}
with
\begin{equation}
\begin{split}
\hdelta(x,s):=\hat{\tau}(x,s)
+
\sum_{s'\in\mathcal S}\{\hat P_{S,1}(s'\cond s)-\hat P_{S,0}(s'\cond s)\}\hat V^*(s').
\end{split}
\end{equation}

To bound the regret of $\hpi$, we first invoke the following performance-difference lemma for average-reward MDPs \citep{even2009online,jin2024feasible}, which expresses the value gap as an expectation under the stationary distribution of a single policy.

\begin{lemm}[Performance Difference Lemma]
Let $Q_\pi(x,s,w)$ be defined as in \eqref{eq:Q_xsw_defi}. Then for any two policies $\pi$ and $\pi'$, the difference between their long-term average outcome is
\begin{equation}
    \mu(\pi')-\mu(\pi) =\EE[\pi']{(\pi'(X,S) - \pi(X,S))\p{Q_\pi(X,S,1)-Q_\pi(X,S,0) } },
\end{equation}
where the expectation is taken over $X\sim P_X, S\sim d_{\pi'}(S)$.
\label{lemm:perf_diff}
\end{lemm}

Recall that for any pair of $(x,s)$, $Q_{\pi^*}(x,s,1)-Q_{\pi^*}(x,s,0) = \delta^*(x,s)$. Thus, Lemma \ref{lemm:perf_diff} implies that
\begin{equation}
\mu(\pi^*) - \mu(\hpi)
=\EE[\hpi]{\p{\pi^*(X,S) - \hpi(X,S)} \delta^*(X,S)}.
\end{equation}
For any fixed $(x,s)$, if $\pi^*(x,s)=1$ and $\hpi(x,s)=0$,
\begin{equation}
0\le \delta^*(x,s)\le \delta^*(x,s) - \hdelta(x,s);
\end{equation}
on the other hand, if $\pi^*(x,s)=0$ and $\hpi(x,s)=1$,
\begin{equation}
0\le -\delta^*(x,s)\le \hdelta(x,s)-\delta^*(x,s).
\end{equation}
Thus,
\begin{equation}
0\le \p{\pi^*(x,s) - \hpi(x,s)} \delta^*(x,s)
\le \abs{\delta^*(x,s) - \hdelta(x,s)},
\end{equation}
and
\begin{equation}
\begin{split}
\mu(\pi^*) - \mu(\hpi)
&\le
\EE[\hpi]{\abs{\delta^*(X,S) - \hdelta(X,S)}}\\
&\le \max_{s\in\mathcal{S}} \Norm{\delta^*(X,s) - \hdelta(X,s)}_{L_1(P_X)}\\
&\lesssim\max_{s\in\mathcal{S}} \Norm{\tau(X,s) - \htau(X,s)}_{L_2(P_X)} + \Norm{\hV^*-V^*}\infty\\
&\qquad\qquad +\max_{w\in\cb{0,1}}\Norm{\hat P_{S,w}-P_{S,w}}_{\infty,1}\Norm{V^*}_\infty.
\end{split}
\end{equation}
By Lemma~\ref{lemm:V_error},
\begin{equation}
\begin{split}
\Norm{\hat V^*-V^*}_\infty
&\lesssim \p{t_0 + t_{\mix}} \p{\Norm{\hat r_0-r_0}_\infty + \max_{s\in\mathcal{S}}\Norm{\htau(\cdot,s)-\tau(\cdot,s)}_{L_2(P_X)} } \\
&\qquad+ M_Y \p{t_0 + t_{\mix}}^2 \max_{w\in\cb{0,1}}\Norm{\hat P_{S,w}-P_{S,w}}_{\infty,1}+\oop\p{\abs{\mathcal{I}_e}^{-1/2}}.
\end{split}
\end{equation}

By Assumption~\ref{assu:stationary}, $\abs{\mathcal{S}}$ is finite. Under the overlap condition in Assumption~\ref{assu:r0_nuisance}, each state-action pair $(s,w)$ is visited with positive stationary probability under the behavior policy. Therefore, by standard root-$n$ convergence of finite-dimensional empirical averages,
\begin{equation}
\begin{split}
\max_{w\in\cb{0,1}}\Norm{\hat P_{S,w}-P_{S,w}}_{\infty,1} = \oop(\abs{\mathcal{I}_t}^{-1/2}).
\end{split}
\end{equation}
For estimating $r_0$, 
define 
\begin{equation}
\bar r_0(s) := \EE{\heta_0(X_i,s)+ \frac{I(W_i=0)}{1-\hpi_0(X_i,s)}\cb{Y_i- \heta_0(X_i,s)} \cond S_i=s, \mathcal{I}_t}.
\end{equation}
Conditionally on the training sample used to estimate $\hat\eta_0(x,s)$ and $\hat\pi_0(x,s)$, 
\begin{equation}
\begin{split}
&\abs{\bar r_0(s)-r_0(s)}\\
&= \abs{\EE{\heta_0(X_i,s)+ \frac{1-\pi_0(X_i,s)}{1-\hpi_0(X_i,s)}\cb{\eta_0(X_i,s)- \heta_0(X_i,s)}  - \eta_0(X_i,s) \cond  S_i=s,\mathcal{I}_t}}\\
&\le \EE{\abs{\frac{\hpi_0(X_i,s)-\pi_0(X_i,s)}{1-\hpi_0(X_i,s)}\cb{\eta_0(X_i,s)- \heta_0(X_i,s)}}\cond  S_i=s,\mathcal{I}_t} \\
&= \oop(\abs{\mathcal{I}_t}^{-1/2})
\end{split}
\end{equation}
by Assumption~\ref{assu:r0_nuisance}. Thus,
\begin{equation}
\begin{split}
\abs{\hat r_0(s) - r_0(s)}
&\le \abs{\hat r_0(s) - \bar r_0(s)} + 
\abs{\bar r_0(s)-r_0(s)}\\
&= \oop(\abs{\mathcal{I}_e}^{-1/2}) + \oop(\abs{\mathcal{I}_t}^{-1/2}),
\end{split}
\end{equation}
where the first term is the empirical fluctuation of the state-$s$ average in~\eqref{eq:baseline_estimate} and is $\oop(\abs{\mathcal{I}_e}^{-1/2})$ since each state has positive limiting frequency and the summands are bounded.

Putting everything together, as $\abs{\mathcal{I}_t}, \abs{\mathcal{I}_e} \asymp n$,
\begin{equation}
\begin{split}
\mu(\pi^*) - \mu(\hpi)
&\lesssim\max_{s\in\mathcal{S}} \Norm{\tau(X,s) - \htau(X,s)}_{L_2(P_X)} + \Norm{\hV^*-V^*}\infty\\
&\qquad\qquad +\max_{w\in\cb{0,1}}\Norm{\hat P_{S,w}-P_{S,w}}_{\infty,1}\Norm{V^*}_\infty\\
&=\oop\p{n^{-(\beta \land \frac{1}{2})}}
\end{split}
\end{equation}

\subsection{Proof of Proposition~\ref{prop:collapsed_state_regret}}

Recall that the oracle policy thresholds the difference-of-Qs at zero:
\begin{equation}
\begin{split}
&\pi^*(x,s) := I\p{\delta^*(x,s)\ge 0 },\\
&\delta^*(x,s):=\eta_1(x,s)-\eta_0(x,s)
+
\sum_{s'\in\mathcal S}\{P_{S,1}(s'\cond s)-P_{S,0}(s'\cond s)\}V^*(s').
\end{split}
\end{equation}
In both cases, we obtain the estimated policy by thresholding an estimated oracle score at zero. 

We start by specializing the exogenous-covariate approach to the tabular setting. 
For each $x\in\mathcal{X}$, $s\in\mathcal{S}$, and $w\in\{0,1\}$, we estimate the conditional mean outcome 
$\eta_w(x,s)$ and the state transition kernels $P_{S,w}(s'\cond s)$ by empirical conditional frequencies, and estimate the marginal law $P_X$ by its empirical distribution. We then form the plug-in CADE estimate
$\htau(x,s)=\heta_1(x,s)-\heta_0(x,s)$, together with a plug-in estimate $\hat r_0(s)$ for $r_0(s)$, and the corresponding plug-in Bellman equation:
\begin{equation}
(\hat T_{S,\text{tab}} v)(s) := \hat r_0(s) + \p{\hat P_{S,0}v}(s) + \sum_{x\in\mathcal{X}} \hat P_X\p{\htau(x,s)+ \p{(\hat P_{S,1}-\hat P_{S,0})v}(s)}_+.
\end{equation}
This removes the additional empirical integration step that motivated the sample-splitting construction in the general case.

By~\eqref{eq:overlap_collapsed}, for each $(x,s)\in\mathcal X\times\mathcal S$ and $w\in\{0,1\}$,
\begin{equation}
\PP{X_i=x,S_i=s,W_i =w}
=\PP{W_i=w\cond X_i=x,S_i=s}\PP{X_i=x,S_i=s}
\ge \frac{\Gamma_1\Gamma_2}{|\mathcal X||\mathcal S|}.
\end{equation}
Therefore, by Hoeffding's inequality,
\begin{equation}
\max_{w\in\{0,1\}}\Norm{\hat{\eta}_w-\eta_w}_\infty
= \oo_p \p{\sqrt{\frac{|\mathcal X||\mathcal S|\log(|\mathcal X||\mathcal S|)}{n}} }.
\end{equation}
Similarly, 
\begin{equation}
\max_{w\in\{0,1\}}\Norm{\hat P_{S,w}-P_{S,w}}_{\infty,1}
= \oo_p \p{|\mathcal S|
\sqrt{\frac{\log(|\mathcal S|)}{n}} }.
\end{equation}

Since the expectation over $X$ in the Bellman operator $\hat T_{S,\text{tab}}$ is now evaluated by finite summation, the additional empirical-integration term in Lemma~\ref{lemm:V_error} disappears.
By Lemma~\ref{lemm:V_error} but replacing $\hat T_{S}$ with $\hat T_{S,\text{tab}}$,
\begin{equation}
\begin{split}
&\Norm{\hat V^*-V^*}_\infty\\
&\qquad\lesssim \p{t_0 + t_{\mix}} \max_{w\in\{0,1\}}\Norm{\hat{\eta}_w-\eta_w}_\infty \\
&\qquad\qquad+ M_Y \p{t_0 + t_{\mix}}^2 \max_{w\in\cb{0,1}}\Norm{\hat P_{S,w}-P_{S,w}}_{\infty,1}\\
&\qquad=\oop\p{\p{t_0 + t_{\mix}}\,\sqrt{\frac{|\mathcal X||\mathcal S|\log(|\mathcal X||\mathcal S|)}{n}}+ \p{t_0 + t_{\mix}}^2\,|\mathcal S|\sqrt{\frac{\log(|\mathcal S|)}{n}}}.
\end{split}
\end{equation}
Thus,
\begin{equation}
\begin{split}
&\Norm{\delta^*-\hdelta}_\infty\\
&\qquad\le \Norm{\hat\tau-\tau}_\infty
+
2\max_w\Norm{\hat P_w-P_w}_{\infty,1}\Norm{V^*}_\infty
+
2\Norm{\hat V^*-V^*}_\infty\\
&\qquad=\oop\p{\p{t_0 + t_{\mix}}\,\sqrt{\frac{|\mathcal X||\mathcal S|\log(|\mathcal X||\mathcal S|)}{n}}+ \p{t_0 + t_{\mix}}^2\,|\mathcal S|\sqrt{\frac{\log(|\mathcal S|)}{n}}},
\end{split}
\end{equation}
and the policy regret is of the same order by the performance difference lemma, Lemma~\ref{lemm:perf_diff},
using a similar argument as in proof of Theorem~\ref{theo:policy_consistency}:
\begin{equation}
\begin{split}
&\mu(\pi^*)-\mu(\hpi)\\
&\qquad\lesssim\Norm{\delta^*-\hdelta}_\infty\\
&\qquad=\oop\p{\p{t_0 + t_{\mix}}\,\sqrt{\frac{|\mathcal X||\mathcal S|\log(|\mathcal X||\mathcal S|)}{n}}+ \p{t_0 + t_{\mix}}^2\,|\mathcal S|\sqrt{\frac{\log(|\mathcal S|)}{n}}}.
\end{split}
\end{equation}

We then consider the collapsed-state benchmark. 
Similarly, in this tabular setting, we can estimate $\eta_w(z)$ and $P_w(z' \cond z)$ by empirical conditional means and transition frequencies.
Let $(\hat\mu^*,\hat V^*)$ solve the empirical Poisson equation with the empirical Bellman operator for the collapsed state variable:
\begin{equation}
(\hat T v)(z) := \max_{w \in \cb{0,1}} 
\p{\heta_w(z) + \sum_{z' \in \mathcal X \times \mathcal S}\hat P_{Z,w}(z'\cond z) v(z')},
\label{eq:poisson_estimator}
\end{equation}
We can then consider the corresponding plug-in policy
\begin{equation}
\begin{split}
&\hat \pi^\col(z) := I\p{\hdelta^\col(z)\ge 0 },\\
&\hdelta^\col(z):=\hat{\eta}_1(z)-\hat{\eta}_0(z)
+
\sum_{z'}\{\hat P_1(z'\cond z)-\hat P_0(z'\cond z)\}\hat V^*(z') .
\end{split}
\end{equation}

This collapsed-state benchmark can be viewed as a special case of the exogenous-covariate formulation by introducing a degenerate covariate $U_i\equiv \star$ and taking the state variable to be $Z_i=(X_i,S_i)$. Then, the collapsed-state Bellman operator
\begin{equation}
(Tv)(z)
:=\max_{w \in \{0,1\}}
\cb{\eta_w(z) + \sum_{z' \in \mathcal X \times \mathcal S} P_w(z' \cond z)v(z')},
\end{equation}
coincides with the state-level Bellman operator in~\eqref{eq:state_bellman} written for the pair $(U_i,Z_i)$, because the expectation over the degenerate variable $U_i$ does nothing, the average baseline $r_0(z)$ becomes $\eta_0(z)$, the CADE $\tau(u,z)$ becomes
$\eta_1(z)-\eta_0(z)$, and the transition kernel is now $P_{Z,w}$. Thus, Lemma~\ref{lemm:V_error} carries over exactly once 
$s$ is replaced by $z$.

By construction, the collapsed-state benchmark is the same Bellman problem as in Lemma~\ref{lemm:V_error}, with state variable $Z=(X,S)$, reward functions $\eta_w(z)$, and transition kernels $P_{Z,w}(\cdot\cond z)$. Thus, from Lemma~\ref{lemm:V_error} (but replacing $\hat T_{S}$ with $\hat T_{Z}$),
\begin{equation}
\begin{split}
\Norm{\hat V^*-V^*}_\infty
&\lesssim \p{t_0 + t_{\mix}} \max_{w\in\cb{0,1}}\Norm{\eta_w-\heta_w}_{\infty}  \\
&\qquad\qquad+ M_Y \p{t_0 + t_{\mix}}^2 \max_{w\in\cb{0,1}}\Norm{\hat P_{Z,w}-P_{Z,w}}_{\infty,1}.
\end{split}
\end{equation}

Again, by~\eqref{eq:overlap_collapsed} and Hoeffding's inequality,
\begin{equation}
\max_{w\in\{0,1\}}\Norm{\hat{\eta}_w-\eta_w}_\infty
= \oo_p \p{\sqrt{\frac{|\mathcal X||\mathcal S|\log(|\mathcal X||\mathcal S|)}{n}} },
\end{equation}
and
\begin{equation}
\max_{w\in\{0,1\}}\Norm{\hat P_{Z,w}-P_{Z,w}}_{\infty,1}
= \oo_p \p{|\mathcal X||\mathcal S|
\sqrt{\frac{\log(|\mathcal X||\mathcal S|)}{n}} }.
\end{equation}
Thus, following the same argument as in proof of Theorem~\ref{theo:policy_consistency},
\begin{equation}
\begin{split}
\mu(\pi^*)-\mu(\hat\pi^\col)
&\lesssim \Norm{\delta^*-\hdelta^\col}_\infty\\
&\lesssim \max_w\Norm{\hat{\eta}_w-\eta_w}_\infty
+
\max_w\Norm{\hat P_{Z,w}-P_{Z,w}}_{\infty,1}\Norm{V^*}_\infty
+
\Norm{\hat V^*-V^*}_\infty\\
&= \oo_p\p{\p{t_0 + t_{\mix}}^2|\mathcal X||\mathcal S|\sqrt{\frac{\log(|\mathcal X||\mathcal S|)}{n}} }
\end{split}
\end{equation}

\section{Proof of Lemmas}

\subsection{Proof of Lemma~\ref{lemm:bellman_equation_state}}

We first write the standard Bellman optimality equation for the Markov decision process whose full state is $Z_i:=(X_i,S_i)$, since our model can be viewed as a Markov decision process with state $Z_i$. 
By the average-reward optimality equation for Markov
decision process \citep{sutton2018reinforcement}, there exist an optimal average reward $\mu^*$ and a relative value function $V_Z^*: \mathcal X \times \mathcal S\to \RR$ such that
\begin{equation}
\mu^* + V_Z^*(z) = \max_{w \in \cb{0,1}} 
\p{\eta_w(z) + \int_{z' \in \mathcal X \times \mathcal S} P_{Z,w}(z'\cond z) V_Z^*(z')},
\label{eq:bellman_equation_collapsed}
\end{equation}
where $P_{Z,w}(z'\cond z)$ denotes the collapsed covariate-state transition kernel.
Under Assumption~\ref{assu:MDP}, $P_{Z,w}(z'\cond z)= P_{S,w}(s'\cond s) P_{X}(x')$, and thus
\begin{equation}
\mu^* + V_Z^*(x,s) = \max_{w \in \cb{0,1}} 
\p{\eta_w(x,s) + \int_{(x',s') \in \mathcal X \times \mathcal S} P_{S,w}(s'\cond s) P_{X}(x') V_Z^*(x',s')}.
\end{equation}
Defining $V^*(s):=\mathbb E_X[V_Z^*(X,s)]$ and taking expectation over $X$ on both sides yield
\begin{equation}
\begin{split}
\mu^* + V^*(s) &= \EE[X]{\max_{w \in \cb{0,1}} 
\p{\eta_w(X,s) + \int_{(x',s') \in \mathcal X \times \mathcal S} P_{S,w}(s'\cond s) P_{X}(x') V_Z^*(x',s')}}\\
&= \EE[X]{\max_{w \in \cb{0,1}} 
\p{\eta_w(X,s) + \sum_{s' \in \mathcal S} P_{S,w}(s'\cond s) V^*(s')}}.
\end{split}
\end{equation}
This proves the first claim of the lemma.

Next, consider any policy $\pi$ that attains the pointwise maximum inside the expectation for each covariate-state pair $(x,s)$. Then,
\begin{equation}
\mu^* + V^*(s) = \EE[X]{
\eta_{\pi(X,s)}(X,s) + \sum_{s' \in \mathcal S} P_{S,\pi(X,s)}(s'\cond s) V^*(s')}.
\end{equation}
Multiplying both sides by the stationary distribution $d_\pi(s)$ and summing over $s\in\mathcal S$, we obtain
\begin{equation}
\begin{split}
&\mu^* + \sum_s d_\pi(s)V^*(s) \\
&\qquad= 
\EE[X]{ \sum_s d_\pi(s)\eta_{\pi(X,s)}(X,s) + \sum_s d_\pi(s)\sum_{s'} P_{S,\pi(X,s)}(s'\cond s) V^*(s')}\\
&\qquad= 
\mu(\pi) + \sum_{s'} d_\pi(s') V^*(s').
\end{split}
\end{equation}
Thus, $\mu^*=\mu(\pi)$ and $\pi$ must be an optimal policy.

\subsection{Proof of Lemma~\ref{lemm:V_error}}
Recall that the state-level operator is defined as
\begin{equation}
\begin{split}
(T_S v)(s) 
&= r_0(s) + P_{S,0}v(s) + \EE[X]{\p{\tau(X,s)+ (P_{S,1}-P_{S,0})v(s)}_+}.
\end{split}
\end{equation}
and empirical version of the operator used in the algorithm is
\begin{equation}
\begin{split}
(\hat T_S v)(s) = \hat r_0(s) + \hat P_{S,0}v(s) + \frac{1}{\abs{\mathcal{I}_e}}\sum_{i\in\mathcal{I}_e}{\p{\htau(X_i,s)+ (\hat P_{S,1}-\hat P_{S,0})v(s)}_+}.
\end{split}
\end{equation}
Then $(\mu^*,V^*)$ and $(\hat \mu^*, \hat V^*)$ satisfy
\begin{equation}
\mu^* \one + V^* = T_S V^*,
\qquad
\hat \mu^* \one + \hat V^* = \hat T_S \hat V^*.
\end{equation}

Define an intermediate Bellman operator 
\begin{equation}
\begin{split}
(\widetilde T_S v)(s) = \hat r_0(s) + \hat P_{S,0}v(s) + \EE[X]{\p{\htau(X,s)+ (\hat P_{S,1}-\hat P_{S,0})v(s)}_+},
\end{split}
\end{equation}
with $(\widetilde \mu^*, \widetilde V^*)$ satisfy
\begin{equation}
\widetilde \mu^* \one + \widetilde V^* = \widetilde T_S \widetilde V^*.
\end{equation}

For any $v : \mathcal S \to \mathbb R$ and any
$s \in \mathcal S$, since $u\mapsto u_+$ is nonexpansive, 
\begin{equation}
\begin{split}
\abs{(\hat T_S v)(s) - (T_S v)(s)}
&\le \Norm{\hat r_0 - r_0}_\infty +\max_s \Norm{\htau(X,s)-\tau(X,s)}_{L_2(P_X)}\\
&\qquad + 3\Norm{v}_\infty\max_{w \in \{0,1\}}\Norm{\hat P_{S,w} - P_{S,w}}_{\infty,1} + \abs{(\hat T_S v)(s) - (\widetilde T_S v)(s)},
\end{split}
\end{equation}
where
\begin{equation}
\begin{split}
\abs{(\hat T_S v)(s) - (\widetilde T_S v)(s)}
= \oop\p{\frac{1}{\sqrt{\abs{\mathcal{I}_e}}}}.
\end{split}
\end{equation}
Thus
\begin{equation}
\begin{split}
\Norm{\hat T_S V^* - T_S V^*}_\infty
&\le
\Norm{\hat r_0 - r_0}_\infty
+\max_s \Norm{\htau(X,s)-\tau(X,s)}_{L_2(P_X)}\\
&\qquad
+3\Norm{V^*}_\infty
\max_{w \in \{0,1\}} \Norm{\hat P_{S,w} - P_{S,w}}_{\infty,1}+\oop\p{\abs{\mathcal{I}_e}^{-1/2}}.
\end{split}
\end{equation}
By Assumption~\ref{assu:uniform_mixing},
for any $s,\widetilde s \in \mathcal S$,
\begin{equation}
\begin{split}
\abs{V^*(s) - V^*(\widetilde s)}
&\le\sum_{t \ge 0}
\abs{\EE[\pi^*]{r_{\pi^*}(S_t)\cond S_0=s}
-\EE[\pi^*]{r_{\pi^*}(S_t)\cond S_0=\widetilde s}}\\
&\le 2 M_Y \sum_{t \ge 0}
\Norm{P_{S,\pi^*}^t(\cdot \cond s) - P_{S,\pi^*}^t(\cdot \cond \widetilde s)}_{\mathrm{TV}} \\
&\le 4 M_Y \p{t_0 + \sum_{t \ge t_0} e^{-t/t_{\mix}}}
\le 8 M_Y \p{t_0 + t_{\mix}}.
\end{split}
\end{equation}
Since $V^*(s_0)=0$,
\begin{equation}
\Norm{V^*}_\infty \le \mathrm{span}(V^*) \le 8 M_Y \p{t_0 + t_{\mix}}.
\end{equation}

Define the oracle and empirical policies
\begin{equation}
\begin{split}
\pi^*(z)&\in\argmax_{w\in\cb{0,1}} \cb{\eta_w(x,s)+(P_{S,w}V^*)(s)},\\
\hat \pi(z)&\in\argmax_{w\in\cb{0,1}} \cb{\heta_w(x,s)+(\hat P_{S,w}\hat V^*)(s)}.
\end{split}
\end{equation}
Then,
\begin{equation}
\mu^*\one + V^* = r_{\pi^*} + P_{S,\pi^*} V^*,\qquad
\hat \mu^*\one + \hat V^* = \hat r_{\hpi} + \hat P_{S,\hpi} \hat V^*,
\label{eq:optimal_equations}
\end{equation}
and thus
\begin{equation}
\begin{split}
\hat \mu^*\one + \hat V^* &= \hat r_{\hpi} + \hat P_{S,\hpi} \hat V^*\\
&= \hat r_{\hpi} + \hat P_{S,\hpi} V^* + \hat P_{S,\hpi}\p{\hat V^*- V^*}\\
&\le \hat T_S V^* + \hat P_{S,\hpi}\p{\hat V^*- V^*},
\end{split}
\end{equation}
because $\hat T_S V^*$ is the pointwise maximization over $w$ evaluated at $ V^*$.
Subtracting $\mu^*\one + V^* = T_S V^*$ gives
\begin{equation}
(\hat \mu^*-\mu^*)\one + \hat V^*- V^*
\le (\hat T_S V^* - T_S V^*) + \hat P_{S,\hpi}\p{\hat V^*- V^*}.
\label{eq:max_to_fixed}
\end{equation}
Iterating,
\begin{equation}
\begin{split}
\hat V^* - V^*
&\le \hat P_{S,\hat\pi}^h(\hat V^* - V^*)
 + \sum_{m=0}^{h-1} \hat P_{S,\hat\pi}^m
 \p{(\hat T_S V^* - T_S V^*) - (\hat\mu^*-\mu^*)\one }.
\end{split}
\end{equation}
Note that at the anchor state $s_0$, $\hat V^*(s_0)=V^*(s_0)=0$. Subtracting $\hat V^*(s_0)-V^*(s_0)$,
\begin{equation}
\begin{split}
\hat V^*(s)-V^*(s)
&\le \sqb{\hat P_{S,\hat\pi}^h(\hat V^*-V^*)}(s)
- \sqb{\hat P_{S,\hat\pi}^h(\hat V^*-V^*)}(s_0)\\
&\qquad
+\sum_{m=0}^{h-1}
\p{\sqb{\hat P_{S,\hat\pi}^m(\hat T_S V^*-T_S V^*)}(s)
-\sqb{\hat P_{S,\hat\pi}^m(\hat T_S V^*-T_S V^*)}(s_0)},
\end{split}
\end{equation}
where we used $\hat P_{S,\hat\pi}^m\one = \one$ to cancel the
term involving $(\hat\mu^*-\mu^*)\one$.
By Assumption~\ref{assu:uniform_mixing},
\begin{equation}
\begin{split}
\hat V^*(s)-V^*(s)
&\le 4\p{t_0+\sum_{m\ge t_0} e^{-m/t_{\mix}}}
\Norm{\hat T_S V^*-T_S V^*}_\infty\\
&\le 8\p{t_0 + t_{\mix}} \Norm{\hat T_S V^*-T_S V^*}_\infty.
\end{split}
\end{equation}
For the other side, \eqref{eq:optimal_equations} implies
\begin{equation}
\begin{split}
\hat \mu^*\one + \hat V^* &= \hat r_{\hpi} + \hat P_{S,\hpi} \hat V^*\\
&\ge \hat r_{\pi^*}+\hat P_{S,\pi^*}\hat V^*\\
&= \hat r_{\pi^*}+\hat P_{S,\pi^*}V^*+\hat P_{S,\pi^*}(\hat V^*-V^*),
\end{split}
\end{equation}
because $\hpi$ is greedy for $\hat T_S$ at $\hat V^*$.
Subtracting $\mu^*\one + V^* = T_S V^*$ gives
\begin{equation}
(\hat \mu^*-\mu^*)\one + \hat V^*- V^*
\ge (\hat r_{\pi^*}-r_{\pi^*})+(\hat P_{S,\pi^*}-P_{S,\pi^*})V^*
+\hat P_{S,\pi^*}(\hat V^*-V^*).
\end{equation}
Iterating,
\begin{equation}
\begin{split}
\hat V^* - V^*
&\ge \hat P_{S,\pi^*}^h(\hat V^* - V^*)
 + \sum_{m=0}^{h-1} \hat P_{S,\pi^*}^m
 \p{(\hat r_{\pi^*}-r_{\pi^*}) + (\hat P_{S,\pi^*}-P_{S,\pi^*})V^* - (\hat\mu^*-\mu^*)\one }.
\end{split}
\end{equation}
Evaluating at $s$ and $s_0$ and subtracting,
\begin{equation}
\begin{split}
\hat V^*(s)-V^*(s)
&\ge \sqb{\hat P_{S,\pi^*}^h(\hat V^*-V^*)}(s)
- \sqb{\hat P_{S,\pi^*}^h(\hat V^*-V^*)}(s_0)\\
&\qquad
+\sum_{m=0}^{h-1}
\sqb{\hat P_{S,\pi^*}^m\p{(\hat r_{\pi^*}-r_{\pi^*}) + (\hat P_{S,\pi^*}-P_{S,\pi^*})V^*}}(s)\\
&\qquad
-\sum_{m=0}^{h-1}\sqb{\hat P_{S,\pi^*}^m\p{(\hat r_{\pi^*}-r_{\pi^*}) + (\hat P_{S,\pi^*}-P_{S,\pi^*})V^*}}(s_0)\\
&\ge
-4\p{t_0+\sum_{m\ge t_0} e^{-m/t_{\mix}}}
\Norm{(\hat r_{\pi^*}-r_{\pi^*}) + (\hat P_{S,\pi^*}-P_{S,\pi^*})V^*}_\infty\\
&\ge
-8 \p{t_0 + t_{\mix}} \p{\Norm{\hat r_{\pi^*} - r_{\pi^*}}_\infty
+
\Norm{\hat P_{S,\pi^*}-P_{S,\pi^*})V^*}_{\infty}}.
\end{split}
\end{equation}
Note that for all $s\in\mathcal{S}$,
\begin{equation}
\hat r_{\pi^*}(s)-r_{\pi^*}(s)
=\hat r_0(s)-r_0(s) + \EE[X]{\pi^*(X,s)\p{\htau(X,s)-\tau(X,s)}}+\oop\p{\abs{\mathcal{I}_e}^{-1/2}}.
\end{equation}
Thus,
\begin{equation}
\Norm{\hat r_{\pi^*}(s)-r_{\pi^*}(s)}
\le\Norm{\hat r_0-r_0}_\infty + \max_s\Norm{\htau(X,s)-\tau(X,s)}_{L_2(P_X)}+\oop\p{\abs{\mathcal{I}_e}^{-1/2}},
\end{equation}
and
\begin{equation}
\begin{split}
\hat V^*(s)-V^*(s)&\ge
-8\p{t_0 + t_{\mix}} \left(\Norm{\hat r_0-r_0}_\infty + \max_s\Norm{\htau(X,s)-\tau(X,s)}_{L_2(P_X)}\right.\\
&\qquad \left. + \Norm{V^*}_\infty\max_{w \in \{0,1\}} \Norm{\hat P_{S,w} - P_{S,w}}_{\infty,1}\right)+\oop\p{\abs{\mathcal{I}_e}^{-1/2}}.
\end{split}
\end{equation}

Putting everything together,
\begin{equation}
\begin{split}
\Norm{\hat V^*-V^*}_\infty
&\lesssim \p{t_0 + t_{\mix}} \p{\Norm{\hat r_0-r_0}_\infty + \max_{s\in\mathcal{S}}\Norm{\htau(X,s)-\tau(X,s)}_{L_2(P_X)} } \\
&\qquad\qquad+ M_Y \p{t_0 + t_{\mix}}^2 \max_{w\in\cb{0,1}}\Norm{\hat P_w-P_w}_{\infty,1}+\oop\p{\abs{\mathcal{I}_e}^{-1/2}}.
\end{split}
\end{equation}

\subsection{Proof of Lemma \ref{lemm:perf_diff}}
It follows directly from Lemma D.8 of \cite{jin2024feasible} that
\begin{equation}
\begin{split}
    \mu(\pi')-\mu(\pi) 
    &=\EE[\pi']{(\pi'(X,S) - \pi(X,S))Q_\pi(X,S,1)}\\
    &\qquad \EE[\pi']{((1-\pi'(X,S)) - (1-\pi(X,S)))Q_\pi(X,S,0) }\\
    &=\EE[\pi']{(\pi'(X,S) - \pi(X,S))\p{Q_\pi(X,S,1)-Q_\pi(X,S,0) } },
\end{split}
\end{equation}
where the expectation is taken over $X\sim P_X, S\sim d_{\pi'}(S)$, as the covariates are generated i.i.d. from $P_X$.

\end{document}